%% file: cusp_amplitudes.tex
\documentclass[reqno,12pt]{article}
\pdfoutput=1
\usepackage{ae}
\usepackage[T1]{fontenc}
\usepackage[ansinew]{inputenc}
\usepackage{mathrsfs}
\usepackage{amsmath}
\usepackage{amssymb}
\usepackage{graphicx}
\usepackage{color}
\definecolor{darkblue}{cmyk}{0.9,0.9,0,0}
\definecolor{darkgreen}{rgb}{0,0.55,0}
\usepackage[colorlinks=true,linkcolor=darkblue,citecolor=darkblue,urlcolor=darkblue]{hyperref}
\usepackage{epsfig}
\usepackage{wasysym}
\usepackage{MnSymbol}

\usepackage{amsmath,amsfonts,amssymb,amsbsy}
\usepackage{graphicx,color}
\usepackage{graphics}
\usepackage{epsfig}
\usepackage{verbatim}
\usepackage{stmaryrd}
\usepackage{epsf}

\usepackage{amsmath, amssymb,graphicx}
\usepackage{epsfig}
\usepackage{verbatim}
\usepackage{amsfonts}
\usepackage{ulem}
\usepackage{cite}
\usepackage{stmaryrd}

\usepackage{tikz}
\usetikzlibrary{arrows,shapes}
\usetikzlibrary{patterns,fadings}

\newcommand{\btp}{\begin{tikzpicture}[baseline=0pt,scale=0.9,line width=0.7pt]}
\newcommand{\btpp}{\begin{tikzpicture}[baseline=-5pt,scale=0.25,line width=0.7pt]}
\newcommand{\etp}{\end{tikzpicture}}

\def\bc{\begin{center}}
\def\ec{\end{center}}

\usepackage{epsf}

\newcommand{\be}{\begin{equation}}
\newcommand{\ee}{\end{equation}}
\newcommand{\ba}{\begin{eqnarray}}
\newcommand{\ea}{\end{eqnarray}}
\newcommand{\nn}{{\nonumber}}

\newcommand{\beaa}{\begin{eqnarray}}
\newcommand{\eeaa}{\end{eqnarray}}

\newcommand{\eps}{\epsilon}

\definecolor{darkgreen}{rgb}{0.1,0.7,0.1}

\newcommand{\Blue}[1]{{\color{blue}#1\color{black}}}

\newcommand{\la}[1]{\label{#1}}

\DeclareFontFamily{OT1}{pzc}{}
\DeclareFontShape{OT1}{pzc}{m}{it}{<-> s * [1.10] pzcmi7t}{}
\DeclareMathAlphabet{\mathpzc}{OT1}{pzc}{m}{it}

\def\({\left(}
\def\){\right)}
\def\[{\left[}
\def\]{\right]}

\def\<{\langle}
\def\>{\rangle}

\def\cO{{\cal O}}

\def\nref#1{(\ref{#1})}

\def\cusp{ {\rm cusp}}

\def\cM{{\cal M}}

\newcommand{\cN}{{\cal N}}

\begin{document}

\thispagestyle{empty}

\renewcommand{\thefootnote}{\fnsymbol{footnote}}
\setcounter{footnote}{0}
\setcounter{figure}{0}
\begin{center}
$$$$
{\Large\textbf{\mathversion{bold}
The cusp anomalous dimension\\ at three loops and beyond
}

\par}

\vspace{1.0cm}

\textrm{Diego Correa$^a$, Johannes Henn$^b$, Juan Maldacena$^b$ and  Amit Sever$^{b,c}$ }
\\ \vspace{1.2cm}
\footnotesize{

\textit{$^{a}$  Instituto de F\'{\i}sica La Plata, Universidad Nacional de La Plata, \\ C.C. 67, 1900 La Plata, Argentina} \\
\texttt{} \\
\vspace{3mm}
\textit{$^{b}$ School of Natural Sciences,\\Institute for Advanced Study, Princeton, NJ 08540, USA.} \\
\texttt{} \\
\vspace{3mm}
\textit{$^c$
Perimeter Institute for Theoretical Physics\\ Waterloo,
Ontario N2J 2W9, Canada} \\
\texttt{}
\vspace{3mm}
}

\par\vspace{1.5cm}

\textbf{Abstract}\vspace{2mm}
\end{center}

\noindent

We derive an analytic formula at three loops for the cusp anomalous
dimension $\Gamma_{\rm cusp}(\phi)$ in $\cN=4$ super Yang-Mills. This is done by exploiting the
relation of the latter to the Regge limit of massive amplitudes.
We comment on the corresponding three loops quark anti-quark potential.
Our result also determines a considerable part of the three-loop cusp anomalous dimension in QCD.
Finally, we consider a limit in which only ladder diagrams contribute to physical observables.
In that limit, a precise agreement with strong coupling is observed.
\vspace*{\fill}

\setcounter{page}{1}
\renewcommand{\thefootnote}{\arabic{footnote}}
\setcounter{footnote}{0}

\newpage
\tableofcontents

\section{Introduction}

The cusp anomalous dimension $\Gamma_{\rm cusp}(\phi)$ was originally
introduced in \cite{PolyakovCusp} as the ultraviolet (UV) divergence of a
Wilson loop with a cusp with Euclidean angle $\phi$. It describes a wide range of interesting physical situations. {One  that will be of special importance in this paper is its relation to}
the infrared (IR) divergences of
massive scattering amplitudes and form factors, see \cite{Korchemsky:1991zp,Becher:2009kw}
and references therein.

In planar $\cN=4$ SYM, using dual conformal symmetry one can argue that
the (Euclidean) Regge limit
\footnote{We call this limit ``Euclidean'' because $s$ and $t$ have the
same sign, as opposed to the usual Regge limit.} $s \gg m^2, t$ of Coulomb branch
amplitudes \cite{Alday:2009zm} is also governed by this function $\Gamma_\text{cusp}$ \cite{Henn:2010bk}.
In this paper, we  use this relation to extract the three-loop value of $\Gamma_{\rm cusp}$ from the scattering amplitudes.

The two-loop contribution to the cusp anomalous dimension in QCD was computed
in ref. \cite{Korchemsky:1987wg} and rederived and simplified in ref. \cite{Kidonakis:2009ev}.
In supersymmetric theories, it is natural to define a generalization of the usual Wilson loop.
As we discuss below, one can define a locally supersymmetric Wilson loop that couples to scalars in addition to gluons. 
The two-loop result for the latter loop in $\cN=4$ SYM is closely related to the result for the standard QCD loop, 
and was obtained in \cite{Makeenko:2006ds,Drukker:2011za}.
We reproduced these formulas as a check of our calculation, see eqs. (\ref{gammacusp1loop}) and (\ref{gammacusp2loop}).

We find the previously unknown three-loop contribution to be given by
 {\allowdisplaybreaks
\begin{align}
\Gamma^{(3)}_{\rm cusp}(\varphi)  \,=&\, - \frac{1}{2} \, \xi \, \left\{  \frac{1}{3}  \varphi \left( \varphi^2  + \pi^2 \right)^2 \right\}
\phantom{...........................................a lot of empty space} \nonumber \\
 & \hspace{-1.5cm} -\frac{1}{2} \, \xi^2 \,
 \Big\{
  -\frac{11}{15} \varphi^5
   -\frac{4}{3} \,{\rm Li}_1(e^{-2\varphi}) \varphi^4
 + \varphi^3  \, \left[
  +\frac{2}{3}  \,{\rm Li}_2(e^{-2\varphi})
     -\frac{20}{3} \zeta _2 \right]
         \nonumber\\
&\hspace{-0.5cm}
+ \varphi^2  \, \left[
 -2 \zeta _3
 -8 \zeta _2\, {\rm Li}_1(e^{-2\varphi})
   +2\, {\rm Li}_3(e^{-2\varphi})
  \right]
         \nonumber\\
&\hspace{-0.5cm}
 +\varphi  \, \left[
 +4 \zeta _2 \,{\rm Li}_2(e^{-2\varphi})
 -25 \zeta _4
    -9  \,{\rm Li}_4(e^{-2\varphi})  \right]
+12
   \,{\rm Li}_5(e^{-2\varphi})   -12 \zeta _5
  \Big\}  \nonumber \\
    & \hspace{-1.5cm}
   - \frac{1}{2} \,  \xi^3  \, \Big\{
  - \frac{2}{5}  \varphi ^5
   - \varphi^3  \left[
   \frac{4}{3}\, {\rm Li}_2(e^{-2\varphi})
   +\frac{8}{3} \zeta _2
   \right]
   +    \varphi^2  \, \left[
      -2
    \, {\rm Li}_3(e^{ - 2 \varphi } ) +2 \zeta _3  \right]
       \nonumber\\
&\hspace{-0.5cm}
      - \varphi  \left[ 4 \, H_{2,2}(e^{-2\varphi})
    +4  \, H_{3,1}(e^{-2\varphi})+4 \zeta _2 \,{\rm Li}_2(e^{-2\varphi})
   +4 \, {\rm Li}_4(e^{-2\varphi})  +6 \zeta _4  \right]
         \nonumber\\
&\hspace{-0.5cm}
     -6\, {\rm Li}_5(e^{-2\varphi})   -4\, H_{2,3}(e^{-2\varphi})-6 \,H_{3,2}(e^{-2\varphi})-6 \,H_{4,1}(e^{-2\varphi})
    +4 \zeta _3
 \,  {\rm Li}_2(e^{-2\varphi})
         \nonumber\\
&\hspace{-0.5cm}
   -2 \zeta _2\, {\rm Li}_3(e^{-2\varphi})+2 \zeta
   _2 \zeta _3+3 \zeta _5
  \Big\} \,. \label{gamma3_intro}
  \end{align}
Here $\xi = \tanh {\varphi}/{2}$, and ${\rm Li}_{n}$ are polylogarithms.
Here $\varphi = i \phi$ is the lorentzian version of the angle (a boost angle).
The $H_{i,j}$ are harmonic polylogarithms \cite{Remiddi:1999ew},
whose definition we recall in section \ref{sec:amplitudes}.
This is one of the main results of this paper.

There are several interesting limits of this function.
\begin{itemize}

\item In the large angle $\varphi\to\infty$ limit, it grows linearly with the angle, $\Gamma_{\rm cusp}(\varphi) \,\propto \, \varphi \Gamma_{\cusp}^{\infty}$.
The coefficient $\Gamma_{\rm cusp}^{\infty}$ is the anomalous dimension of a light-like, or null, Wilson loop. It is also related to
the high-spin limit of anomalous dimensions of composite operators
\cite{Korchemsky:1988si,Korchemsky:1992xv,Alday:2007mf}. 
It is determined exactly by an integral equation \cite{BES}.

\item Small angle limit. At $\phi=0$, we have the straight line, which is $1/2$ BPS, and its loop corrections vanish. The order $\phi^2$ correction to the BPS configuration is
related to the energy loss of an accelerated quark.
We have $ \Gamma_{\rm cusp}(\phi)  = - \phi^2 \, B(\lambda, N) + \cO(\phi^4)$, where the ``Bremstrahlung function'' $B(\lambda,N)$
is exactly known \cite{Correa:2012at,Fiol:2012sg}. In \cite{Correa:2012at,Fiol:2012sg} this
function was also related to  other
observables, such as the power radiated by a moving quark,
the two point function of the displacement operator on the Wilson loop and the stress tensor
expectation value in the presence of a Wilson line.

\item $\Gamma_{cusp}(\phi)$ gives the quark anti-quark potential on $S^3$ for a configuration
which is separated by an angle $\delta = \pi - \phi$, see figure \ref{CuspDiagram}(b).
 The limit $ \delta \to 0 $ gives the quark anti-quark potential potential in flat space. That limit is considered in section \ref{LadderResummation}.
\end{itemize}

In ${\cal N}=4$ super Yang Mills we can also introduce a second angle at a cusp \cite{Drukker:1999zq}. This second angle is related to the fact that the locally supersymmetric Wilson loop observable contains a coupling to a scalar. This coupling selects a direction $\vec n$, where $\vec n$ is a point on $S^5$. The Wilson loop operator is given by \cite{Maldacena:1998im}
\be
 W \sim Tr[ P  e^{ i \oint A\cdot dx + \oint |dx| \vec n\cdot \vec \Phi } ]
  \ee
where we wrote it in Euclidean signature. One can consider a loop with a constant direction $\vec n $, with $\partial_\tau \vec n =0$. Consider such straight Wilson line making a sudden turn by an angle $\phi$, see figure \ref{CuspDiagram}(a). At the cusp we could consider the possibility of changing the direction $\vec n$ by an angle $\theta$, $\cos \theta = \vec n \cdot\vec n'$, where $\vec n$ and $\vec n'$ are the directions before and after the cusp. In that case the Wilson loop develops a logarithmic divergence of the form
\be \la{cuspdef}
 \langle W \rangle \sim e^{ - \Gamma_{\cusp}(\phi,\theta) \log{\Lambda_\text{UV}\over\Lambda_\text{IR} } }
\ee
where $\Lambda_\text{IR/UV}$ are the IR and UV cutoff energies, respectively.
Thus, we have a cusp anomalous dimension, $\Gamma_{cusp}(\phi, \theta)$, which is a function of two angles $\phi$ and $\theta$. The former is the obvious geometric angle and the latter is an internal angle.
\begin{figure}[h]
\centering
\def\svgwidth{14cm}
 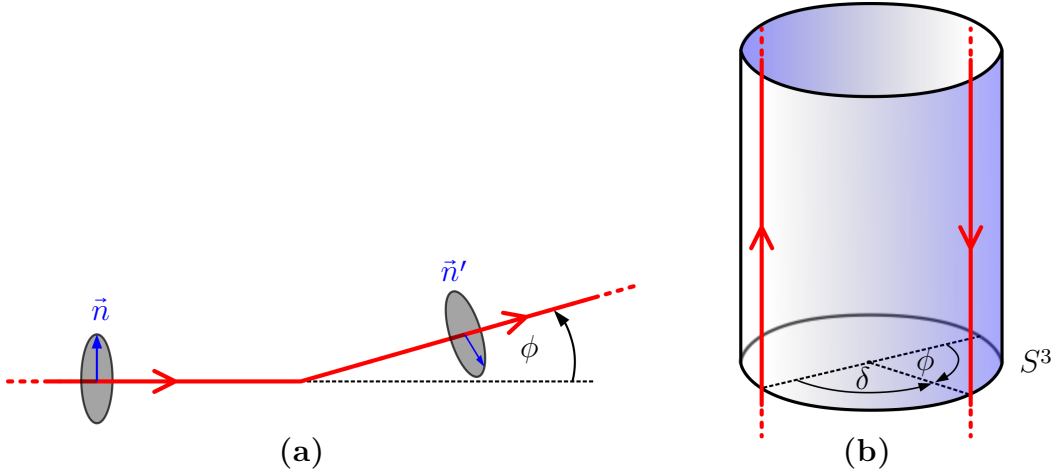
\caption{({\bf a}) A Wilson line that makes a turn by an angle $\phi$ in Euclidean space.
The vectors $\vec n$, $\vec n'$ denote the direction that sets the coupling to the scalar.
({\bf b}) Under the plane to cylinder map,  the same line is mapped to a quark anti-quark configuration. The quark and antiquark are sitting at two points  on $S^3$ at a relative angle of $\delta = \pi - \phi$.
Of course, they are extended along the time direction.}\label{CuspDiagram}
\end{figure}

The same generalized cusp anomalous dimension $\Gamma_{cusp}(\phi, \theta)$ also characterizes the planar IR divergences that arise when scattering massive W bosons on the Coulomb branch of ${\cal N}=4$ SYM. There, $\cos\theta=\vec n_1\cdot\vec n_2$ is the angle between the Coulomb branch expectation values $\<\vec\Phi\>=\text{diag}(h_1\vec n_1,h_2\vec n_2,\dots)$ associated to a pair of color adjacent external W bosons $W_{1,i}$ and $W_{i,2}$. This generalized cusp
anomalous dimension was computed to leading and subleading order in weak and strong coupling in refs. \cite{Drukker:1999zq} and \cite{Drukker:2011za}, respectively. We expect that,
 at three loops,
 $\Gamma_{\rm cusp}(\varphi, \theta)$ can be obtained from the result at $\theta = 0$ of eq. (\ref{gamma3_intro}) by setting $\xi=(\cosh \varphi - \cos \theta)/\sinh \varphi$.

Ladder diagrams give the ``most complicated'' contribution at three loops. That is the $\xi^3$ piece in eq. (\ref{gamma3_intro}).
More generally, we find that up to three loops ladder diagrams are the only contribution to the $\xi^{L}$ term where $L$ is the number of loops. We expect that to be true to all loop orders. One can therefore wonder if there is a physical quantity that is computed by summing only ladder diagrams. We find that this is indeed the case! Having a second angle $\theta$, there are new limits one can take. In particular one can consider the limit where $i\theta\to\infty$, $\lambda\to0$ with $\hat\lambda=\lambda\, e^{i\theta}/4$ (and therefore also $\lambda^L \xi^L$) held fixed. That limit select the scalar ladders diagrams and is the subject of section \ref{sec:ladders}. The sum of these ladder diagrams can be performed by finding the ground
state of a one dimensional Schr\"odinger problem.  It is possible to take the large $\hat \lambda$
limit. We  compare this with the strong coupling answer in the large $e^{i\theta}$ limit and
we find that they agree. In principle, it did not have to agree, since the order of limits is
different. However, as in the BMN \cite{Berenstein:2002jq} or large charge limits \cite{Kruczenski:2003gt}, the two limits commute and we get
a precise agreement. For a particular angle, namely $\phi =0$, we can solve it completely as
a function of the coupling.

There are two appendices. In appendix \ref{QCDintegrals}
we comment on direct relations between integrals appearing in massive form
factors and the Wilson loop under consideration. We explicitly give a sum of two three-loop
Wilson loop diagrams that contributes directly to the three-loop form factor in QCD.
We also compute a certain type of Wilson loop integral with interaction vertices (or equivalently, form factor integral)
analytically to all loop orders. We observe that up to a simple factor, the answer at $L$ loops is a homogeneous
polynomial of degree $2L-1$ in the cusp angle $\varphi$ and $\pi$.
Finally, appendix \ref{appendix-integrals} contains the result for the Regge limit of individual four-point integrals that
contribute to $\Gamma_{\rm cusp}$.

\section{$\Gamma_{\rm cusp}(\lambda,\varphi)$ from scattering amplitudes}
\label{sec:amplitudes}

\subsection{Regge limit and soft divergences in massive $\cN=4$ super Yang-Mills}

\begin{figure}[t]
\begin{center}
\includegraphics[width=0.85\textwidth]{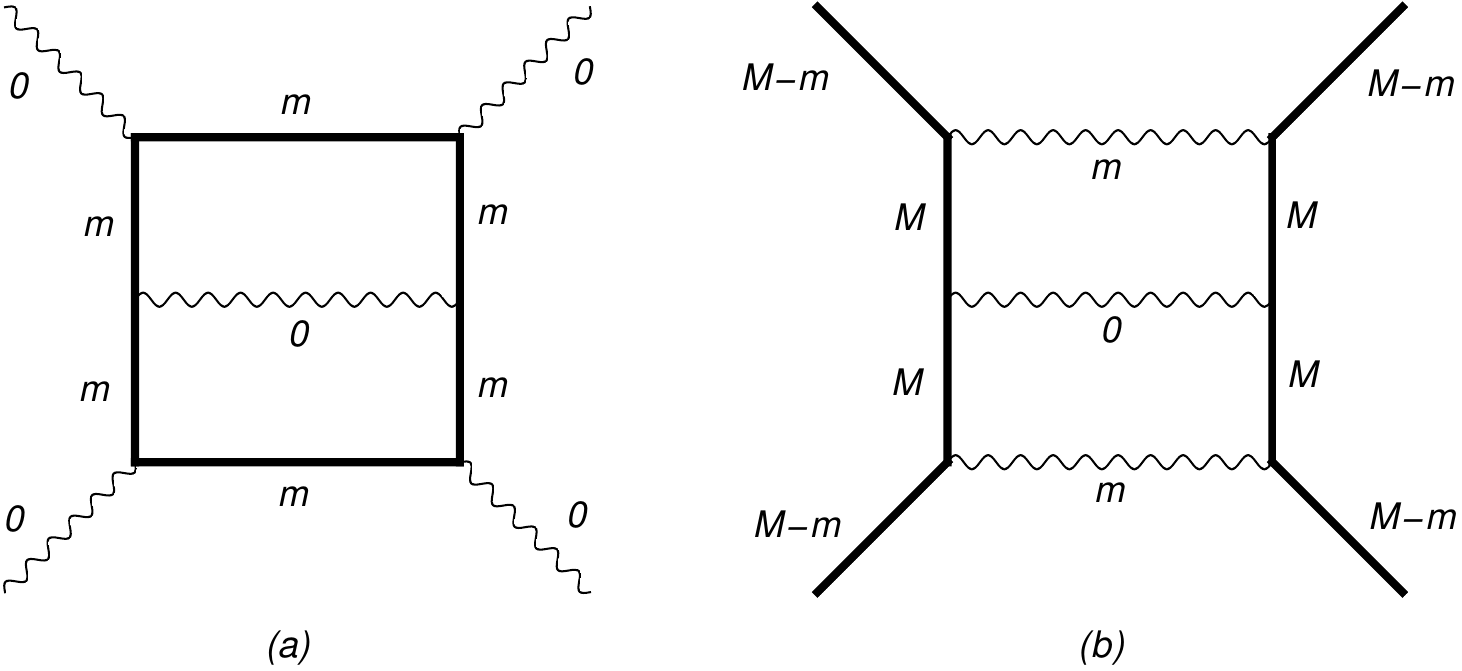}
\caption{Sample two-loop diagram contributing to the four-particle amplitude.
Solid and wavy lines denote massive and (almost) massless particles, respectively.
The precise masses are given by the labels.
Dual conformal symmetry implies that the same function $\cM(u,v)$ describes two different physical situations:
The Regge limit $s\to \infty$ of (a) is equivalent to the Bhabha-type scattering (b), where the outer wavy lines
have a small mass that regulates the soft divergences.}
\label{fig:regge-soft}
\end{center}
\end{figure}

The cusp anomalous dimension  $\Gamma_{\rm cusp}(\varphi)$ appears in (supersymmetric)
Wilson loops with (Minkowskian) cusp angle $\varphi$. It can also be obtained from a leading
infrared (soft) divergence of amplitudes or form factors involving massive particles \cite{Korchemsky:1991zp}.

In particular, it can be extracted from massive amplitudes on the Coulomb branch of $\cN=4$ SYM \cite{Henn:2010bk}.
These amplitudes are obtained by giving a vacuum expectation value to some of the scalars
of $\cN=4$ SYM. The string theory dual of this setup \cite{Berkovits:2008ic}
 suggests that the amplitudes defined
in this way have an exact dual conformal symmetry \cite{Alday:2009zm}.
Consider the four-scalar amplitude $\cM$ with on-shell conditions $p_{i}^2 = - (h_{i} - h_{i+1})^2$,
where $h_{i+4} \equiv h_{i}$. Here $h_i$ are four nonzero
eigenvalues of the Higgs fields, which we take
to all point in the same direction.
The Mandelstam variables are $s=(p_{1}+p_{2})^2$ and $t=(p_{2}+p_{3})^2$.
\footnote{We follow the $-+++$ metric conventions of ref. \cite{Alday:2009zm}, so that $s$ is negative for
positive center of mass energy. The amplitude will be real for $s$ and $t$ both positive.}
A priori the amplitude could depend on five dimensionless ratios built from the
Poincar\'e invariants $s,t,h^2_{i}$.
Dual conformal symmetry implies that it is a function of two variables
only,
\begin{align}
\cM(s, t, h_{1}, h_{2}, h_{3}, h_{4} ) = \cM(u,v)\,,
\end{align}
where
\begin{align}
 u=\frac{ h_1\, h_3}{s+(h_1-h_3)^2}\,,\qquad v=\frac{ h_2 \, h_4}{t+(h_2-h_4)^2}\,.
\end{align}
$u$ and $v$ are invariant under dual conformal transformations.
$\cM(u,v)$ is known explicitly at one loop, where it is given in terms of logarithms and dilogarithms. It is an interesting open
question what its functional dependence is at higher loops. In this paper, we determine the $u \to 0$ limit up to
three loops.

Following ref. \cite{Henn:2010bk}, there are two mass configurations that are of special interest two us:
\begin{itemize}
\item[(a)] The equal mass configuration $h_{i} = m$, which implies that the external states are massless, $p_{i}^2=0$,
and we have $u=m^2 / s\,, v=m^2 /t$. In the limit $u,v \ll 1$, with  $u/v$ fixed one obtains the (IR-divergent) massless amplitudes
at the origin of the Coulomb branch. In particular, in the four-point case discussed here, the latter are known to all loop orders.
Here $m$ acts as an IR regulator.
\item[(b)] The two-mass configuration $h_{1}=h_{3}=m$, $h_{2}=h_{4}=M$, where the on-shell conditions are $p_{i}^2 = -(M-m)^2$,
and $u=m^2/s\,, v=M^2/t$. In the limit of $M>>m$, the kinematical configuration is that of Bhabha scattering. Here we have external
massive particles of mass $M$, and $m$ acts as an IR cutoff if we keep $s$ and $t$ of order $M^2$.
In string theory, the different masses come from strings ending on stacks of D3 branes at different radial coordinates in AdS space.
See ref. \cite{Alday:2009zm} for more
details. In the planar limit we can view this as the scattering of  a massive quarks and a massive antiquark.
\end{itemize}
Thanks to dual conformal symmetry, both (a) and (b) are described by the same function $\cM(u,v)$.

We will be interested in the limit $u\ll1$ of $\cM(u,v)$ \cite{Henn:2010bk}.
This limit has different physical interpretations for cases (a) and (b).
The two configurations (in the limit) are illustrated in Fig.~\ref{fig:regge-soft}.
In the interpretation (a) of $\cM(u,v)$, this is the Regge limit $s \to \infty$.
In (b), in the limit we have heavy particles of mass $M-m \approx M$ that interact by exchanging light particles of small mass $m$.
The mass of the light particles regulates soft divergences.
This is similar in spirit to giving the photon a small mass in QED.
In the planar limit, and from the point of view of the low energy degrees of freedom, we
are scattering a massive quark and anti-quark pair. Thus there are only two regions that
give rise to soft divergences\footnote{If we were scattering massive adjoint particles there would be four regions that give rise to soft divergences.}.  See Fig.~\ref{fig:regge-soft}(b).
The overall soft IR divergence can be obtained in the
eikonal approximation. This leads naturally to a cusped Wilson loop \cite{Korchemsky:1985xj}. 
One can then identify the coefficient of the
IR divergence of the amplitude with (minus) the coefficient of the UV divergence of the cusped Wilson loop.
Therefore we expect \cite{Henn:2010bk}
\footnote{Note that we changed the sign and normalization in the definition of $\Gamma_{\rm cusp}(\lambda,\varphi)$
w.r.t. ref. \cite{Henn:2010bk}. We have $\Gamma^{ there}_{\rm cusp}(\lambda,\varphi) = -2 \Gamma^{ here}_{\rm cusp}(\lambda,\varphi)$.
\label{footnote_conventions}}
\begin{align}\label{amplitude-regge}
\log\left( \cM(u,v) \right) \;\; \underset{u\to 0}{\longrightarrow}  \;\; (\log u ) \,\Gamma_{\rm cusp}(\lambda,\varphi) + \cO(u^0) \,.
\end{align}
Note that there are two soft regions of the amplitude that contribute.
These regions correspond to either the upper or lower propagator of mass $m$ in Fig.~\ref{fig:regge-soft}(b)
having soft momentum.

We have the perturbative expansion
\begin{align}
\Gamma_{\rm cusp}(\lambda,\varphi)   = \sum_{L\ge1} \, \left( \frac{\lambda}{8 \pi^2} \right)^L \, \Gamma^{(L)}_{\rm cusp}(\varphi)  \,,\qquad \lambda= {g_{\rm YM}^2 N}\,.
\end{align}
The logarithmic contribution on the r.h.s. of (\ref{amplitude-regge}) comes from soft exchanges between the heavy
particles. In this soft region, the amplitude can be approximated by form factors.
In this limit, the Minkowskian cusp angle is a natural variable.
It is defined as
\begin{align}
\cosh \varphi =  \frac{p_{2} \cdot p_{3}}{ \sqrt{ (p_{2})^2  (p_{3})^2 } } \,,
\end{align}
where $p_{2}$ and $p_{3}$ are the (ingoing) momenta forming the cusp.
The relation to the Euclidean cusp angle is $\phi = i \varphi$.
Using the definition $v = M^2/(p_{2}+p_{3})^2$ and the on-shell conditions
$p_{2}^2 = p_{3}^2 = -M^2$, we have
\begin{align} \label{defvarphi}
\cosh \varphi  = 1+\frac{1}{2 v} \,.
\end{align}
Another natural variable, which is also used frequently in the literature e.g. on
Bhabha scattering integrals, is
\begin{align}
x = \frac{\sqrt{1+4 \,v} -1 }{\sqrt{1+4 \,v}+1}=e^{-\varphi}
\end{align}
For $0<v<\infty$ (i.e. in the Euclidean region) $x$ is real and ranges from $0$ to $1$.
The inverse relation to $v$ is
\begin{align}
v = \frac{x}{(1-x)^2}={1\over4\sinh^2{\varphi\over2}} \,.
\end{align}
Notice that this relation is invariant under $x \to 1/x$.
\footnote{Note that in the physical region, where $v$ has a small imaginary part coming from the Feynman propagator
prescription, the sign of this imaginary part has to be included in the $x \to 1/x$ transformation.}
This symmetry corresponds to $\varphi \to -\varphi$.
In fact, we will find that our results, which are functions of $v$, are invariant under this
transformation.\footnote{
The fact that $\Gamma_{cusp}$ should be analytic around $x=1$ or $\phi =0$ is very clear when
we interpret it as a quark-antiquark potential on the sphere, see figure \ref{CuspDiagram}.}

\subsection{Evaluation of the four-particle amplitude to three loops}

\begin{figure}[t]
\begin{center}
\includegraphics[width=0.9\textwidth]{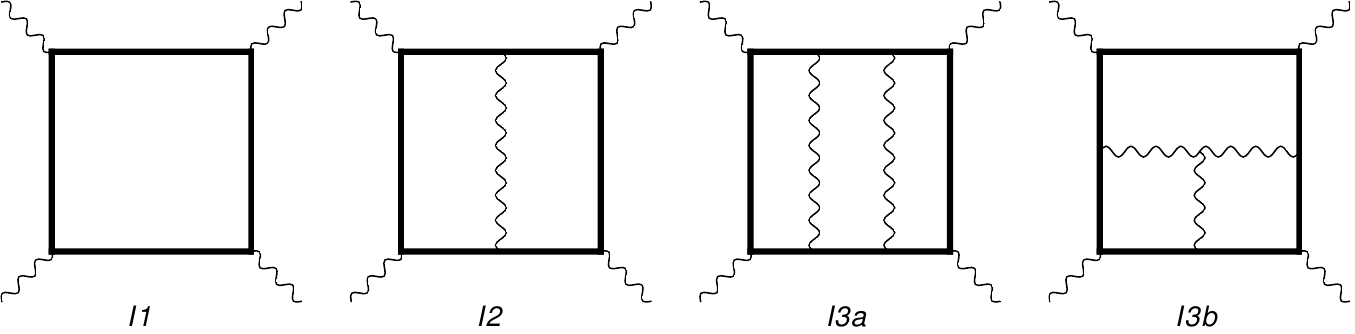}
\caption{Integrals contributing to the four-particle amplitude to three loops. Solid and wavy lines denote massive and massless propagators, respectively. Overall normalizations of $s$ and $t$, as well as a loop-momentum dependent numerator factor in $I_{3b}$ are not displayed.}
\label{fig:integrals}
\end{center}
\end{figure}

The four-particle amplitude on the Coulomb branch has the perturbative expansion\begin{align}\label{Mpert}
\cM = 1+ \sum_{L \ge 1} \left( \frac{\lambda}{8 \pi^2} \right)^L \, \cM^{(L)}(u,v)\,.
\end{align}
To three loops, it is given by  \cite{Bern:1997nh,Henn:2010bk,Bern:2010qa}
\begin{align}
\cM^{(1)}(u,v) =& -\frac{1}{2} I_{1}(u,v) \,, \label{M1} \\
\cM^{(2)}(u,v) =& +  \frac{1}{4}  \left[ I_{2}(u,v)  + I_{2}(v,u) \right] \,, \label{M2} \\
\cM^{(3)}(u,v) =& -\frac{1}{8} \left[ I_{3a}(u,v) + I_{3a}(v,u) + 2 I_{3b}(u,v) + 2 I_{3b}(v,u) \right]  \label{M3} \,.
\end{align}
Here $I_{1}, I_{2}, I_{3a}, I_{3b}$ are massive scalar four-point integrals.
They are shown in Fig~\ref{fig:integrals}, and their precise definition is given
in ref. \cite{Henn:2010bk}, where they have been evaluated in various limits.
Note that the expressions in eq. (\ref{M1})-(\ref{M3}) are valid for generic $u,v$, i.e. for finite values of the mass(es) \cite{Bern:2010qa}.

We are interested in the Regge limit $u \ll 1$.
Note that the rate of divergence of the scalar integrals without numerators
can be predicted a priori. It follows from their topology and can be determined by the counting procedure
of ref. \cite{Eden}.
For example, the one- and two-loop integrals have the following small $u$ expansions
\begin{align}
I_{1}(u,v) \,\; \underset{u\to 0}{\longrightarrow} \,\; & \log u \, I_{1,1}(v)  + I_{1,0}(v) + \cO(u) \,, \\
I_{2}(u,v) \,\; \underset{u\to 0}{\longrightarrow} \,\; & \log u \, I_{2,1}(v) + I_{2,0}(v) + \cO(u) \,,\\
I_{2}(v,u)  \,\; \underset{u\to 0}{\longrightarrow} \,\; & \log^2 u \, I^{r}_{2,2}(v) + \log u \, I^{r}_{2,1}(v) + I^{r}_{2,0}(v) + \cO(u) \,.
\end{align}
(Here the superscript $r$ indicates that the expansion terms in the last line
come from $I_{2}$ rotated by 90 degrees, i.e. with $u$ and $v$ exchanged.)
For the integral $I_{3b}$, which has a loop-momentum dependent numerator, the counting rules do
not apply. Its Regge behavior was analyzed in ref. \cite{Henn:2010bk}.

We wish to determine the coefficients of the logarithms in the Regge limit.
In order to compute $\Gamma_{\rm cusp}$ at three loops, we need all lower-loop
coefficient functions (including the $\cO(u^0)$) ones, and the $\log u$ terms at
three loops. Note that all $\cO(u)$ terms can be safely dropped. This
is an advantage over dimensional regularization.

The four-particle integrals in the Regge limit are very closely
related to form factor integrals. One could in principle use the method
of differential equations \cite{Gehrmann:1999as} in order to compute the latter.
Here, we use a shortcut by making a convenient ansatz for the result.
In order to do that, it is helpful to have an idea of the kind of functions that
can appear in general.
The solutions to the type of differential equations mentioned above can typically be
written in terms of harmonic polylogarithms (HPLs) \cite{Remiddi:1999ew}
of argument $x$.
These functions are defined recursively by integrating the following
kernels,
\begin{align}
f_{1}(x) = \frac{1}{1-x} \,,\qquad f_{0}(x) = \frac{1}{x} \,,\qquad f_{-1}(x)=\frac{1}{1+x} \,.
\end{align}
The starting point for the recursion are the weight one functions
\begin{align}
H_{1}(x) = -\log(1-x) \,,\qquad H_{0}(x) = \log(x) \,,\qquad H_{-1}(x)=\log(1+x) \,.
\end{align}
Higher weight HPLs are defined in the following way,
\begin{align}
H_{a_1, a_2 , \ldots, a_n}(x) = \int\limits_{0}^{x} \, f_{a_1}(t) \, H_{a_2 , \ldots, a_n}(t) \,dt  \,.
\end{align}
The subscript of $H$ is called the weight vector.
We use an abbreviation common in the literature.
If $m$ zeros are to the left of $\pm 1$, they are removed and
$\pm 1$ replaced by $\pm (m+1)$. For example,
\begin{align}
H_{0,0,1,0,-1}(x) = H_{3,-2}(x)\,.
\end{align}
In related studies of massive vertex-type integrals, this set
of functions was sufficient to describe all occurring integrals,
see e.g. \cite{Czakon:2004wm,Anastasiou:2006hc,Gluza:2009yy}.

Motivated by this fact, we wrote down an ansatz in terms of harmonic polylogarithms
for each integral, and determined the coefficients in this ansatz
by evaluating various limits, for $x \to 0$, or $x\to 1$, and by comparing
against generic data points obtained from numerical integration.
In doing so, we found the Mathematica implementation of harmonic
polylogarithms \cite{Maitre:2005uu} and the Mellin-Barnes tools \cite{Czakon:2005rk}
useful. Following this procedure, we were able to find analytical formulas for
all coefficient functions needed to compute $\Gamma_{\rm cusp}$ at
three loops.

As an example, at one loop, we have.
\begin{align}
I_{1,1} =& \,
\xi \, 2 \log x \,, \\
I_{1,0} =& \, \xi \,   \left[ - \pi^2 -4 H_{1,0}(x) - 4 H_{-1,0}(x) \right] \,.\label{intI10}
\end{align}
Here we introduced the useful abbreviation $\xi = (1-x)/(1+x)$.
The formulas are valid for $0<x<1$ and can be analytically continued to other regions, as we describe below.

The results for the two- and three-loop integrals are given in Appendix \ref{appendix-integrals}.
We now proceed to present the result for $\Gamma_{\rm cusp}$.

\subsection{Analytic three-loop result for $\Gamma_{\rm cusp}$}

The perturbative expansion of the cusp anomalous dimension is
\begin{align}\label{def_pert_gammacusp}
\Gamma_{\rm cusp}(\lambda, \varphi) = \sum_{L \ge 1} \left( \frac{\lambda}{8 \pi^2} \right)^L \, \Gamma^{(L)}_{\rm cusp}(\varphi) \,,   \,,\qquad \lambda= {g_{\rm YM}^2 N}\,,
\end{align}
where $g_{\rm YM}$ is the Yang-Mills coupling, and $N$ the number of colors.
As was already mentioned, the result up to two-loop results was known.

As explained above, we compute $\Gamma^{(3)}_{\rm cusp}(\varphi)$ by evaluating the
(Euclidean) Regge limit of the four-particle amplitude, thanks to eq. (\ref{amplitude-regge}).
The Regge limit of all integrals contributing to eq. (\ref{Mpert}) can be found in
Appendix \ref{appendix-integrals}.

We find the following results to three loops, valid in the Euclidean region $x>0$,
{\allowdisplaybreaks
\begin{align}
\Gamma^{(1)}_{\rm cusp}(\varphi)  \,=&\, - \frac{1}{2}\, \xi  \left[2  \log x \right] \,,   \label{gammacusp1loop} \\
\Gamma^{(2)}_{\rm cusp}(\varphi)  \,=&\, - \frac{1}{2}\, \xi \left[ -\frac{2}{3}   \log x \left( \log^2 x + \pi^2 \right)  \right]  \nonumber \\
& \,  \,- \frac{1}{2}\, \xi^2 \left[ \frac{2}{3} \log^3 x +2 \log x \left( \zeta_2 + {\rm Li}_{2}(x^2) \right)- 2 {\rm Li}_{3}(x^2) +2 \zeta_{3} \right]\,,  \label{gammacusp2loop}\\
\Gamma^{(3)}_{\rm cusp}(\varphi)  \,=&\, - \frac{1}{2}\, \xi \left[  \frac{1}{3}  \log x \left( \log^2 x + \pi^2 \right)^2 \right] \nonumber \\
 & \hspace{-1.5cm} - \frac{1}{2}\, \xi^2
 \Big[
 8 \zeta _2 H_{-2,0}(x)-4 \zeta _3 H_{0,0}(x)-8 \zeta _2 H_{2,0}(x)+16 \zeta _2
   H_{-1,0,0}(x)
  -40 \zeta _2 H_{0,0,0}(x)
\nonumber\\
&\hspace{-0.5cm} -16 \zeta _2 H_{1,0,0}(x)+32 H_{-4,0}(x)-32
   H_{4,0}(x)
  +24 H_{-2,0,0,0}(x)-24 H_{2,0,0,0}(x)
  \nonumber \\
  &\hspace{-0.5cm}+32 H_{-1,0,0,0,0}(x)-88
   H_{0,0,0,0,0}(x)-32 H_{1,0,0,0,0}(x)-25 \zeta _4 H_0(x)-12 \zeta _5
  \Big] \nonumber \\
 &\hspace{-1.5cm} - \frac{1}{8}\, \xi^3 \Big[
 16 \zeta _3 H_{0,0}(x)-32 \zeta _2 H_{-2,0}(x)+32 \zeta _2 H_{2,0}(x)+64 \zeta _2 H_{0,0,0}(x)-128
   H_{-4,0}(x)  \nonumber  \\
&\hspace{-0.5cm}    +128 H_{4,0}(x)+64 H_{-3,-1,0}(x)-64 H_{-3,0,0}(x)-64 H_{-3,1,0}(x)+64 H_{-2,-2,0}(x) \nonumber \\
&\hspace{-0.5cm}  -64 H_{-2,2,0}(x)  -64 H_{2,-2,0}(x)+64 H_{2,2,0}(x)-64 H_{3,-1,0}(x)+64 H_{3,0,0}(x) \nonumber \\
 &\hspace{-0.5cm}   +64 H_{3,1,0}(x)-64
   H_{-2,0,0,0}(x)
    +64 H_{2,0,0,0}(x)+192 H_{0,0,0,0,0}(x)-32 \zeta _3 H_{-2}(x) \nonumber \\
    &\hspace{-0.5cm}  +32 \zeta _3 H_2(x)  -32 \zeta _2
   H_{-3}(x)+24 \zeta _4 H_0(x)+32 \zeta _2 H_3(x)+8 \zeta _2 \zeta _3+12 \zeta _5
  \Big] \,.   \label{gammacusp3loop}
\end{align}
}
The one- and two-loop results were known and confirm those
quoted in the literature \cite{Makeenko:2006ds,Drukker:2011za}, see also \cite{Korchemsky:1987wg,Kidonakis:2009ev}.
The three-loop result is new.

Using simple relations between harmonic polylogarithms of argument $x^2$ and $x$, and their product algebra (see e.g. \cite{Maitre:2005uu}),
we can rewrite the $\xi^2$ and $\xi^3$ terms at three loops in a simpler way.
Moreover, replacing the HPLs $H_{n}(x^2)$ appearing in the $\xi^2$ term by classical polylogarithms
${\rm Li}_{n}(x^2)$, as at two loops, leads to eq. (\ref{gamma3_intro}) given in the introduction.
This is the main result of this section.

Let us discuss the branch cut structure of $\Gamma_{\rm cusp}$.
The formulas presented above are valid in the Euclidean region, $0<x<1$.
They are manifestly real for $0<x<1$.
Some of the functions have branch cuts along
$x\in [-\infty, 0]$ or $x \in [1,\infty] $.
Recall that we expect the results to have the symmetry $x\to1/x$.
Therefore the latter branch cuts should be spurious.
One may check this in the case of polylogarithms using the identity
\begin{align}
{\rm Im}\left( {\rm Li}_{n}(x+i 0 ) \right) = \frac{\pi}{(n-1)!}  \log^{n-1} x  \,,\qquad x>1\,,
\end{align}
where $i 0$ indicates that $x$ has an infinitesimally small positive imaginary part.
In general, one can easily verify this property, as well as the $x\to 1/x$ symmetry, using
relations between HPLs of argument $x$ and $1/x$  \cite{Remiddi:1999ew,Maitre:2005uu}.

\begin{figure}[t]
\begin{center}
\includegraphics[width=0.65\textwidth]{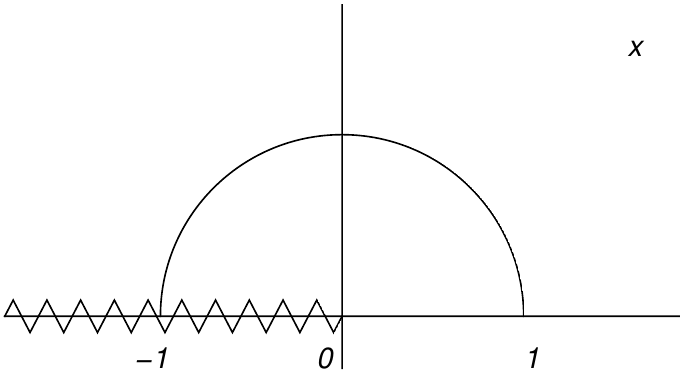}
\caption{Analytic structure of $\Gamma_{\rm cusp}$. The Euclidean region is $x>0$.
Below threshold, $x$ is a phase, $x=e^{i \phi}$. Above threshold, we have that $-1<x<0$, with $x$ having
an infinitesimal positive imaginary part. The zigzag line denotes branch cuts along $[-\infty,0].$}
\label{fig:complex}
\end{center}
\end{figure}

In the physical region, we have to remember the $i0$ prescription of the Feynman propagators.
The sign of $i0$ depends on whether we discuss the Wilson loop, or the scattering amplitudes (and
it depends on the metric conventions.) We will take $x$ to have a small positive imaginary part.
There is a threshold which corresponds to the creation of two massive on-shell particles at $x=-1$
(or, equivalently, $v=-1/4$, i.e. $t=-4 M^2$),
which naturally divides the physical region into two domains, above threshold and below threshold.
Below threshold, $x$ lies on the unit circle in the complex plane, $x=\exp(i \phi)$,
whereas above threshold $x$ is real and satisfies $-1<x<0$.
The analytic continuation to these physical regions is discussed in \cite{Anastasiou:2006hc}.
We summarize the analytic structure for complex $x$ in Fig.~\ref{fig:complex}.

We stress that the final result for $\Gamma_{\rm cusp}$ is much simpler than the individual contributions
presented in Appendix \ref{appendix-integrals}.
We already mentioned that there are different ways of writing the final result.
More interesting than what specific functions are used to express the result in might be the
question whether its symbol \cite{symbols1,symbols2} has any special features.
In fact our ansatz that the result could be expressed just in terms of harmonic polylogarithms
implies that only the entries $1\pm x$, $x$ can appear in the symbol. 
Inspecting the above results, 
we see that the symbol can be written in a way such that its entries are either $x$ or $(1-x^2)/x$.
\footnote{The fact that the symbol can be written just in terms of the entries $x$ and $(1-x^2)/x$ implies that
it has a symmetry under $x \to -x$. Note however that this is not a symmetry of the function. In particular,
the latter acquires an imaginary part for $x<0$, while it is real for $x>0$.}
This property is reflected in the fact that we only needed harmonic polylogarithms with indices $0$ and $1$
(when using $x^2$ as argument). It is also remarkable that the index $1$ appears at most twice.
Finally, we remark that the first entry of the symbol is always $x$, as required by the branch cut
structure of the loop integrals.

Let us now consider various interesting limits of $\Gamma_{\rm cusp}$.
\begin{itemize}
\item In the large $\varphi$ limit (small $v$ limit), $\Gamma_{\rm cusp}(\lambda, \varphi)$ is linear in $\varphi$,
and the coefficient of $\varphi$ is the anomalous dimension of a light-like Wilson loop \cite{Korchemsky:1987wg}.
We reproduce the result for illustration, and as a check of our conventions\footnote{See footnote \ref{footnote_conventions}. We introduced a
factor of $1/2$ on the RHS of eq. (\ref{largecusp}) in order to have the same definition for $\Gamma^\infty_{\rm cusp}(\lambda)$ as in ref. \cite{Henn:2010bk}, and elsewhere in the literature.},
\begin{align}\label{largecusp}
\lim_{\varphi \to \infty} \,  \Gamma_{\rm cusp}(\lambda, \varphi)
= \frac{1}{2} \, \varphi \, \Gamma^{\infty}_{\rm cusp}(\lambda)  + \cO(\varphi^0) \,,
\end{align}
where
\begin{align}
\Gamma^{\infty}_{\rm cusp}(\lambda)  =
2 \, \left(\frac{\lambda}{8 \pi^2}\right) - 2 \zeta_2 \, \left(\frac{\lambda}{8 \pi^2}\right)^2 + 11 \zeta_{4} \, \left(\frac{\lambda}{8 \pi^2}\right)^3 + \cO(\lambda^4)  \,.
\end{align}
These results are correctly reproduced by eqs. (\ref{gammacusp1loop}) - (\ref{gammacusp3loop}).
(In order to take the limit on the three-loop result, the simpler form given in eq. (\ref{gamma3_intro})
is convenient, since its logarithmic dependence as $x\to 0$ is already manifest.)
Moreover, since our result is exact in $v$, sub-leading terms in this limit can also be obtained.
In fact, based on the discussion in \cite{Henn:2010bk}, we might expect that a stronger version
of eq. (\ref{largecusp}) should hold, namely
\begin{align}\label{largecusp2}
\lim_{\varphi \to \infty}\, \Gamma_{\rm cusp}(\varphi)  =\frac{1}{2} \, \varphi \, \Gamma^{\infty}_{\rm cusp}(\lambda) +  \, \tilde{\mathcal G}_{0}(\lambda) +\cO(e^{-\varphi} ) \,,
\end{align}
where
\begin{align}\label{gamma-collinear}
\tilde{\mathcal G}_{0}(\lambda) = -\zeta_{3} \,  \left(\frac{\lambda}{8 \pi^2}\right)^2 + \left( \frac{9}{2} \zeta_{5} - \zeta_2 \zeta_3 \right) \,  \left(\frac{\lambda}{8 \pi^2}\right)^3 + \cO(\lambda^4)\,,
\end{align}
is the collinear anomalous dimension in Higgs regularization.
Equation (\ref{largecusp2}) can be justified if one can show that the limits $\lim_{v\ll1} \lim_{u\ll1}$
and $\lim_{v\ll1} \lim_{u,v\ll1\,, u/v \; {\rm fixed}}$ commute for $M(u,v)$. It was observed in ref. \cite{Henn:2010bk}
that to three loops, these limits {\it do not} commute in general for the individual loop integrals contributing to $M(u,v)$,
however they {\it do} commute for $M(u,v)$ to that order.

Assuming that eq. (\ref{largecusp2}) holds true, it is interesting that it relates the collinear anomalous dimension $\tilde{\mathcal G}_{0}(a)$
for massive amplitudes to the cusp anomalous dimension.

\item In the opposite limit, small $\varphi$ (large $v$), we find
\begin{align}
\Gamma_{\rm cusp}^{(3)}(\varphi)
= & \frac{15}{2} \,\zeta_{4} \, \varphi^2 + \cO\left(\varphi^4 \right) \,.
\end{align}
Taking into account eq. (\ref{defvarphi}) we see that in the limit $1/v \approx  \varphi^2$.
Hence we have, for the leading terms as $\varphi\to 0$,
\begin{align}
\Gamma_{\rm cusp}(\lambda,\varphi) =  \varphi^2 \left[ \frac{1}{2}  \left(\frac{\lambda}{8 \pi^2}\right) -   \frac{\pi^2}{6} \,  \left(\frac{\lambda}{8 \pi^2}\right)^2 +  \frac{\pi^4}{12} \, \left(\frac{\lambda}{8 \pi^2}\right)^3 + \cO(\lambda^4) \right] + \cO(\varphi^4) \,.
\end{align}
This expansion is in perfect agreement with the three loop expansion of the exact result in
\cite{Correa:2012at,Fiol:2012sg}.

\item The limit $x\to -1$. The point $x=1$ corresponds to the
threshold $v=-4 M^2$ of creating two massive particles. It corresponds to $\delta = \pi -\phi \to 0$, which gives the quark antiquark potential in flat space,
see figure \ref{CuspDiagram}. In other words, when the quark and antiquark are very close to
each other on the sphere, their potential is the same as the one they would have in flat space.

In order to reach this limit, we first have to analytically continue to the physical region,
where $x<0$ (and has a small imaginary part coming from the Feynman prescription).
We observe that the terms multiplying $\xi, \xi^2, \xi^3$ in $\Gamma_{\rm cusp}$ vanish as $x \to -1$
(reached after suitable analytic continuation, as discussed above).

For one and two loops, we verified the limit given in \cite{Drukker:2011za}.
It is, in our notation (recall that $x = e^{i \phi}\,, \phi= i \varphi$, $\delta = \pi -\phi$),
\begin{align}
\Gamma_{\rm cusp}^{(1)}(\phi \to \pi) \, =& \, - \frac{2 \pi}{\delta} + \ldots \,,\la{oneloop} \\
\Gamma_{\rm cusp}^{(2)}(\phi \to \pi) \, =& \,  -\frac{8 \pi}{\delta } \log\left( \frac{ 2 \delta }{e} \right)+ \ldots \,. \la{twoloops}
\end{align}
At three loops, we find
\begin{align}\la{threeloops}
\Gamma^{(3)}_{\rm cusp}(\phi \to \pi) \,=& \, - \frac{8}{3} \pi^4 \frac{1}{\delta^2}
- \frac{1}{\delta} \bigg[
16 \pi  \log ^2\left(\frac{2 \delta }{e}\right)+\frac{16}{3} \pi ^3
   \log \left(\frac{2 \delta }{e}\right)  \nonumber \\
 &  +16 \pi  \log \left(\frac{2
   \delta }{e}\right)+36 \pi  \zeta_3 +\frac{4 \pi ^3}{3}-24 \pi \bigg] + \ldots
\end{align}
Most of these contributions come from the $\xi^3$ terms, though the $\xi^2$ term contributes
with a ${ 1 \over \delta} $ and ${ \log \delta \over \delta }$ terms that have been already
included in \nref{threeloops}. The term of order $\xi^1$ does not contribute in this limit.
The logs, as well as the $1/\delta^2$ seem to be in contradiction with the naive expectation. We will discuss the interpretation of this result  in section \ref{LadderResummation}.

\end{itemize}

\section{ A limit that selects the ladder diagrams }\la{sec:ladders}

As we observed in explicit perturbative computations, the term with the highest power of $\xi$ seems
to be given by ladder diagrams up to three loops. One can wonder if this pattern continues to all loops.
Here we argue that it does and that there is a particular limit of $\Gamma_{cusp}(\phi, \theta)$ that
isolates such terms.

Notice that such terms contain the highest powers of $\cos \theta$. This highest power of
$\cos \theta $ can only come from a diagram where there are $L$ scalars ending on each Wilson lines
at the $L$th loop order. The only such diagrams are ladder diagrams.
We conclude that if we take the scaling limit
\be \la{scalinglimit}
  \lambda \to 0 ~,~~~~~~~~~~~~~ e^{ i \theta } \to \infty ~,~~~~~~~~~{\rm with } ~~~~ \hat \lambda =
{ \lambda e^{ i \theta } \over 4 } = {\rm fixed }
\ee
then the ladder diagrams are the only contribution to $\Gamma_{cusp} \longrightarrow \Gamma_{lad}(\hat \lambda , \phi)$.
Note that after the scaling limit we get a non-trivial function of $\phi$ and $\hat \lambda$.
In this section we derive a Schr\"odinger problem whose solution is the sum of ladder diagrams.
We also solve it exactly for $\phi =0$, and for general angles at strong coupling. Notice
that in this limit the coupling to the scalars takes the form
$  Z e^{\pm  i \theta/2} + \bar Z e^{ \mp i \theta/2} $, where we have ``$+$'' for one of the lines and
``$-$'' for the other line.
For the first line, in  the scaling limit, we get
\be \la{Zcoupling}
 \sqrt{\lambda } { 1 \over 2}  \int dt (  Z e^{ i \theta/2} + \bar Z e^{ - i \theta/2} )  \to
 \sqrt{ \hat \lambda } \int dt  Z
 \ee
  For the other line, the only surviving term is the coupling to $\bar Z$. In this limit all bulk interactions vanish and we have the free theory. So the problem is the same
as having a free complex matrix field $Z$ and computing the expectation value in
the presence of the source \nref{Zcoupling}, and the corresponding one for $\bar Z$.

\subsection{Ladder diagrams and the Schr\"odinger problem }

It is convenient to think about the problem on the sphere.
Then the scalar propagators are
\be
 { \lambda \over 8 \pi^2 } { \cos \theta \over 2 ( \cosh (\tau - \sigma ) + \cos \phi ) } ~~\longrightarrow~~~
P(\tau, \sigma)  = { \hat \lambda \over 8 \pi^2 } { 1 \over[ \cosh(\tau - \sigma) + \cos \phi ] }
\ee
where we wrote the propagator in the scaling limit \nref{scalinglimit}. Here $\tau$ and $\sigma$ are
the global time coordinates of the two endpoints.

As explained in\cite{Erickson:1999qv}
 the sum of ladder diagrams can be performed by first introducing a quantity
$ F( T , S)$ which is defined as the sum over all planar ladder diagrams, where we
integrate all insertions up to $\tau \leq T$ and $\sigma \leq S$. Then the derivatives of $F$ are
\be \la{laddeq}
\partial_T \partial_S F(T,S) = P(T,S) F(T,S)
\ee
Then we write $x = T-S$ and $ y = ( T+S)/2$ and we make an ansatz of the form
$F = \sum_n e^{ - \Omega_n y } \Psi_n(x) $. Then eq. \nref{laddeq}
becomes
\be \la{schpro}
\[ - \partial_x^2-  { \hat \lambda \over 8 \pi^2 } { 1 \over ( \cosh x + \cos \phi ) }\] \psi(x) =
- { \Omega^2 \over 4 } \psi(x) = 2 E_{Sch} \psi (x)
\ee
In principle we could find all the eigenvalues of this problem. Note however, that we are only
interested in the long time behavior of the sum, which is governed by  the lowest eigenvalue, $\Omega_0$. We then get $Z \sim e^{  \Omega_0 ({\rm Time } ) } $. The sign of
$\Omega_0$ is not obviously fixed by \nref{schpro}, but we have fixed it here, with $\Omega_0>0$,
so that get the expected sign at weak coupling, for example.
Note that the Schr\"odinger problem, \nref{schpro}, has a discrete set of bound states with negative
Schr\"odinger energy\footnote{Since the potential is negative, there is always at least one  bound state.}, $E_{Sch}$,
(real $\Omega$) and a continuum with $E_{Sch}>0$, or $\Omega$ imaginary.

 We remark that for given $\hat{\lambda}$ and $\phi$, it is possible to compute the
 ground state energy numerically.

\subsection{Exact solution for $\phi =0$}

In the particular case of $\phi =0$, the potential in eq. \nref{schpro}
becomes the P\"oschl-Teller potential $\cosh^{-2}(x/2)$ \cite{Poschl:1933zz}.
In this case the Schr\"odinger problem can be solved exactly by
a variety of techniques.
By the change of variables
\be
 z = { 1 \over 1 + e^{ x } }
 \ee
one can map equation \nref{schpro} to a hypergeometric equation,
 \be\la{exactenergy}
 \psi = [ z ( 1- z )]^{\Omega/2} F(  \Omega - \Omega_0  , \Omega + \Omega_0 + 1 , 1 + \Omega) ~,~~~~~ \Omega_0  = { 1 \over 2} \left[ - 1 + \sqrt{ { \hat \lambda  \over \pi^2 } +  1} \right]
\ee
 Imposing the decaying boundary conditions at $x = \pm \infty $ we see that the possible
 eigenvalues for the bound states  are
 \be
  \Omega_n = \Omega_0 - n ~,~~~~~~~~~~~~~  0 \leq n < \Omega_0
  \ee
  where $n$ is an integer. This gives a finite number of bound states, which depends on
the coupling.    This is in contrast to
 the sum over ladders for the anti-parallel lines in flat space where the number of
  bound states is infinite beyond a certain coupling \cite{Klebanov:2006jj}.
  In particular, the ground state wave function is particularly simple
  \be \la{gsta}
  \psi_0 = { 1 \over ( \cosh{ x/2} )^{\Omega_0 } }
  \ee

 We can also easily compute the first correction away from the $\phi=0$ limit by expanding
 the potential in \nref{schpro}. The first term is proportional to $\phi^2/\cosh^4(x/2)$.
 We can sandwich it between the ground state wave function \nref{gsta} to get the first
 correction
 \be
   - { 2 \Omega_0 \delta \Omega_0 \over 4 } = - { \hat \lambda \over 4 \pi^2 } \phi^2
   { \langle \psi_0 | { 1 \over ( 2 \cosh^2 x/2  )^2 } |\psi_0 \rangle \over
    \langle \psi_0 |   |\psi_0 \rangle }  = - { \hat \lambda \over 4 \pi^2 } \phi^2
    \frac{\Omega_0  (\Omega_0 +1)}{4 (4 \Omega_0  (\Omega_0 +2)+3)}
\ee
Here we used the following formula in order to compute the expectation value,
\begin{align}
 { \langle \psi_0 | { 1 \over ( 2 \cosh^2 x/2  )^2 } |\psi_0 \rangle \over
    \langle \psi_0 |   |\psi_0 \rangle }  = \frac{f(1)}{f(-1)} \,,\qquad f(a) = \int\limits_0^1 dz \left[ z (1-z) \right]^{\Omega_{0} + a}
\end{align}
Then the sum over ladders, up to order $\phi^2$ is
\be \label{full_phi2}
\Gamma^{lad} = { 1 - \sqrt{ \kappa +1 } \over 2 }   -   { \phi^2 \over 16 } \kappa \left( { 1 + \sqrt{ 1 + \kappa } \over 1 + \kappa + 2 \sqrt{\kappa + 1 } } \right) + \cO(\phi^4) ~, ~~~~~~~~~ \kappa = { \hat \lambda \over \pi^2 }
\ee

One could also compute the $\phi^4$ term by using second order perturbation
theory.

We can use eq. (\ref{full_phi2}) as a consistency check of our perturbative calculation of the ladder diagrams.
The sum of the ladders at $L$ loops is given by the $\xi^L$ term of $\Gamma^{(L)}_{\rm cusp}$ that was computed
in section \ref{sec:amplitudes}. Expanding those terms for small angle, we find
\begin{align}
\Gamma^{lad} = \kappa \left[ -\frac{1}{4} - \frac{\phi^2}{24} + \cO(\phi^4)\right] +
\kappa^2 \left[ \frac{1}{16} + \frac{5 \phi^2}{288} + \cO(\phi^4)\right] +
\kappa^3 \left[ -\frac{1}{32} - \frac{43 \phi^2}{3456} + \cO(\phi^4)\right] + \cO(\kappa^4) \,.
\end{align}
This is in perfect agreement with the small $\kappa$ expansion of eq. (\ref{full_phi2}).

\subsection{Perturbative solution in $\hat \lambda$}\la{perturbativesphere}

Here we show how to solve the Schr\"odinger equation for any angle perturbatively in the coupling $\hat{\lambda}$.

To leading order at weak coupling we can approximate the potential by a delta function, since the energy
is very small,
\be \label{deltapotential}
{ 1 \over \cosh x + \cosh \varphi } \sim  { 2 \varphi \over \sinh \varphi } \delta(x)
\ee
which makes sure that we get the right result at first order in the coupling.

To obtain the solution  at higher orders, we use the following procedure.
It is convenient to perform the change of variables
\begin{align}
\Psi(x) = \eta(x) e^{-{\Omega_0}\,x /{2} }
\end{align}
At leading order, we have $\eta(x) = 1$, in agreement with eq. (\ref{deltapotential}).

Since the exponential factor is the correct solution as $x \to \infty$, we can set  the boundary condition
$\eta(\infty) = 1$. This normalizes the solution, but also is stating that we do not have the
growing solution so that we have picked a unique solution of the equations. So, when we
integrate the equation all the way to $x=0$ we will find an $\Omega_0$ dependent value for
the first derivative at the origin.
We note that we can determine $\Omega_0$ from $\eta$ thanks to the boundary condition
\begin{align}
\partial_{x} \Psi(x)|_{x=0} = 0 \quad \longrightarrow \quad 2 \, \partial_x \log \eta(x) |_{x=0}= \Omega_0
\end{align}
which follows from the $x \to -x$ symmetry of the problem and that the ground state wavefunction
is symmetric.
Defining a new variable $w = e^{-x}$ we have
\begin{align}\label{eqeta}
  \partial_w w \partial_w \eta = - \Omega_0   \partial_w \eta + \hat \kappa     \left[ { 1 \over w + e^\varphi} - { 1 \over w + e^{ -\varphi} } \right] \eta  ~,~~~~~~~~
\hat \kappa = { \hat \lambda \over 8  \pi^2 \sinh \varphi }
\end{align}
In this form, it is clear that the differential equation has four regular singular points,
$w=0,\infty, -e^\varphi , - e^{-\varphi}$, so that it is a particular instance of the Heun
equation.
Expanding $\Omega_0 = \hat\kappa \Omega_0^{(1)} + \hat\kappa^2 \Omega_0^{(2)} + \ldots$ and $\eta = 1 + \hat\kappa \eta^{(1)} + \ldots$.
At order $\hat\lambda$, eq. (\ref{eqeta}) becomes
\begin{align}
 \partial_w w \partial_w\eta^{(1)}(w) =     \left[ { 1 \over w + e^\varphi} - { 1 \over w + e^{ -\varphi} } \right]
\end{align}
This equation is easily integrated
\begin{align}
\eta^{(1)} =   \,  \int\limits^{w}_{0} \frac{dw'}{w'} \int\limits_0^{w'} \, dw'' \, \left[ \frac{1}{w'' + e^{\varphi}} - \frac{1}{w'' + e^{-\varphi}} \right]
\end{align}
where we used the boundary conditions at $w=0$ (or $x = \infty$).
In fact, in order to determine $\Omega^{(1)}_0$, we only need to carry out the first integration.
We find
\begin{align}
\Omega^{(1)}_0 =  - 2 \partial_w \eta^{(1)}|_{w=1} =
 2  \varphi
\end{align}
Remembering that the contribution to the ladders is given by $\Gamma^{lad} = - \Omega_0$, we find
perfect agreement with eq. (\ref{gammacusp1loop}), in the limit $i \theta \longrightarrow \infty$ \nref{scalinglimit}.

One can see that at any loop order $L$, the result for $\Omega_0^{(L)}$ can be written as a $(2L-1)$-fold
iterated integral, multiplied by a factor of $( \hat \lambda / \sinh \varphi )^L$.

\subsection{Comparison with strong coupling}\la{strongcoupling}

At strong coupling the potential becomes very deep and we can approximate the energy
by simply the minimum of the potential, at $x=0$.  This then gives
\be \la{ladstrong}
\Gamma^{lad} = - \Omega_0 = - { \sqrt{ \hat \lambda } \over 2 \pi \cos{ \phi \over 2 } } ~,~~~~~~
{\rm for }~~~ \hat \lambda \gg 1
\ee
We have verified that this is in agreement with the strong coupling computation of ref. \cite{Drukker:2011za}.
Namely, \cite{Drukker:2011za} find that the cusp anomalous dimension at strong coupling has
the form
\be \la{drugen}
\Gamma_{cusp} \sim \sqrt{\lambda } F(\phi, \theta )
\ee
If we now take $i \theta \to \infty$, we find that
\be \la{drula}
F(\phi, \theta) \propto { e^{ i \theta/2} \over \cos{\phi \over 2 }}
\ee
We see that inserting \nref{drula} into \nref{drugen} we get \nref{ladstrong}.
In order to see this from the formulas given in \cite{Drukker:2011za},
one sets $q= i r$ with real $r$, and expands for small $p$. In this way one can check
that the coefficient matches precisely with \nref{ladstrong}.

This matching is a bit surprising.
The ladders limit, $\lambda\to0$ with $\hat\lambda$ fixed, is different from the strong coupling limit $\lambda\to\infty$ with $i\theta\gg1$ fixed. That is, the result could in principle depend on the order in which the limits $i \theta \to \infty$ and $\hat\lambda \to \infty$ are taken. A heuristic explanation for this match is the following.
 As we discussed around \nref{Zcoupling}, in the large $e^{i\theta}$ limit one of the Wilson
 lines is sourcing mainly the $Z$ field and the other the $\bar Z$ field.
 Thus the configuration can be viewed as many $Z$ fields with a low density of derivatives.
These derivatives build up  the Wilson loop.

The situation is therefore similar to the BMN limit \cite{Berenstein:2002jq}, or more precisely the classical large charge limit discussed in \cite{Kruczenski:2003gt}.
In that limit a similar match with strong coupling is found for the first two orders in the effective coupling. Here we have checked the leading order term. In principle, the formulas
in \cite{Drukker:2011za} should also allow us to compute the subleading term.
From the Schr\"odinger problem point of view, it is trivial to compute the subleading term, one
simply has to consider the harmonic oscillator approximation around the minimum of the potential
to get
\be
  \la{ladstrongc}
\Gamma^{lad} = - \Omega_0 = - { ({ \sqrt{ \hat \lambda }  - 2 \pi } ) \over 2 \pi \cos{ \phi \over 2 } } + \cO(1/\sqrt{\hat \lambda} )  ~,~~~~~~
{\rm for }~~~ \hat \lambda \gg 1
\ee
and further orders could be computed by straightforward perturbation theory.

\section{ The anti parallel lines limit }
\label{LadderResummation}

When $\delta  \equiv \pi - \phi \to 0$ we have a quark and anti-quark on the sphere separated
by a very small distance. In this limit we expect to find the quark-antiquark potential in
flat space. However, the approach to the limit is tricky because it does not commute with
perturbation theory. Namely, if we first expand for small $\delta$ for fixed $\lambda$ we get
\be \label{smalldelta}
\Gamma_{cusp}(\phi , \theta, \lambda) \sim  -{\alpha_{\rm flat}(\theta , \lambda)  \over \delta } + \cO(\delta^0 )
\ee
where $\alpha_{\rm flat}$ is the coefficient for a quark antiquark potential in flat space for
two straight lines. That is,
\be
V_\text{flat}(r,\lambda,\theta)=-{\alpha_\text{flat}(\theta , \lambda)\over r}=-\lim_{T\to\infty}{1\over T}\log\<W_\talloblong\>
\ee
where $W_\talloblong$ is a rectangular Wilson loop of separation $r$ along a time $T$.
On the other hand if we first expand in $\lambda$ and then, for each fixed order in $\lambda$,
we go to small $\delta$, we do not find the naively expected behavior \nref{smalldelta}.
The reason is that there is a mixing between the color degrees of freedom of the quark-antiquark
pair and low energy degrees of freedom in the bulk, namely bulk modes of energy of order
$\lambda/r$. This causes IR divergences in perturbation theory starting at two loops
in ${\cal N}=4$ SYM  \cite{Erickson:1999qv,Pineda:2007kz}
and at four loops in QCD \cite{Appelquist:1977es}.\footnote{Note that in the QCD literature
on the quark antiquark potential, the loop order is sometimes defined differently.}
The origin of this is the following. A quark-antiquark
pair in the singlet color combination can emit a massless adjoint and then be left in the adjoint.
The force between the quarks in the singlet
and adjoint channel are different.
 In the planar approximation, the force in the adjoint is zero.
For example,
 a long time exchange of a scalar would contribute to the potential $V_\text{flat}$ an expression of the form\footnote{The $-1$ subtraction in (\ref{longdis}) is a counter term in the effective IR theory canceling a power UV divergence.}
\vspace{-0.5cm}
\be \la{longdis}
{\includegraphics[scale=0.25]{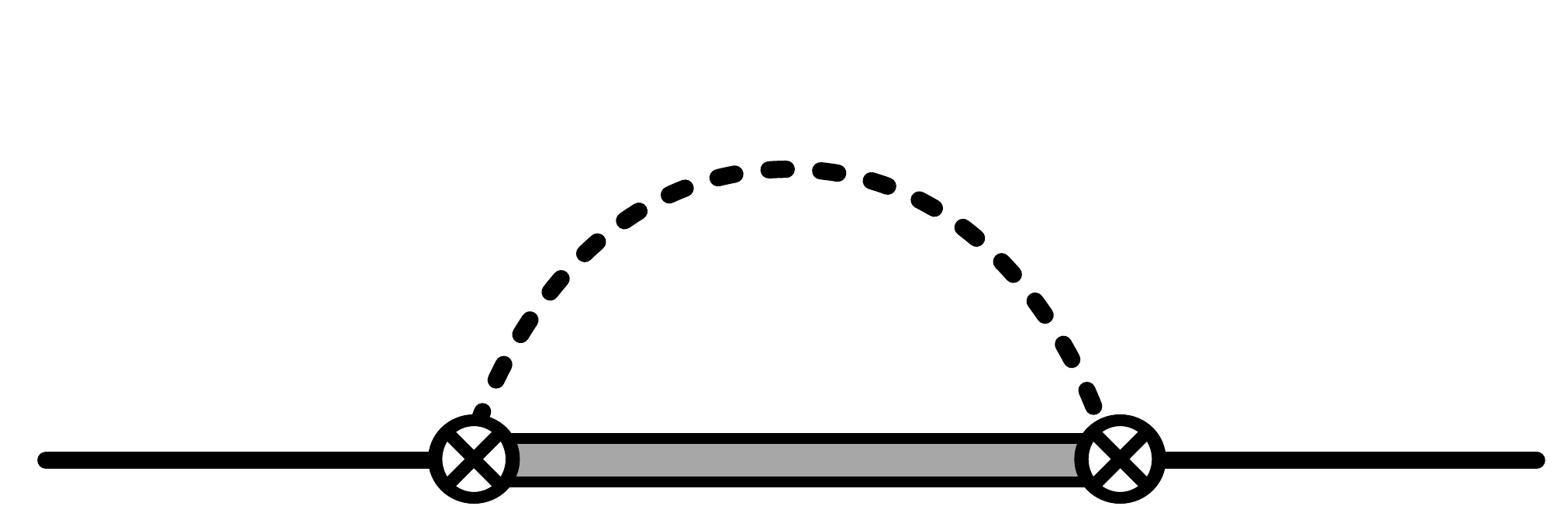}}\quad =\quad { \lambda \over2 \pi^2 } \int\limits_{\epsilon_{UV}}^\infty { d t \over t^2 } ( e^{-t V_\text{singlet} } - 1)
\ee
 where $V_\text{singlet} = { \lambda \over 4 \pi r }+\cO(\lambda^2) $ is minus the singlet potential. This is the binding
 energy we loose when we emit an adjoint, in the planar approximation. Here $1/t^2$ is the large $t$ approximation to the propagator. The UV cutoff of the effective IR theory, $\epsilon_{UV}$, is of the order of the separation $r$.

This integral is perfectly finite in the IR. However, if we first expand in $\lambda$, then  we see that the term
 of order $\lambda^2$ has a logarithmic IR divergence, the term of order $\lambda^3$ has a power
 law IR divergence, etc. If we are on the sphere, we should replace \nref{longdis} by
 \vspace{-0.5cm}
\be \la{spherelong}
{\includegraphics[scale=0.25]{eft1loop.pdf}}\quad =\quad  { \lambda \over 2 \pi^2 } \int\limits_{\epsilon_{UV}}^\infty { d t \over 2 (\cosh t -1) } ( e^{ - t V_{sing} } - 1)
\ee
where now $V_\text{singlet}={ \lambda \over 4 \pi\delta}+\cO(\lambda^2)$ and $\epsilon_{UV}$ is of order $\delta$. If we first expand the integrand in powers of $\lambda$ we now get integrals that are perfectly IR convergent. The long time cutoff is just the size of the sphere that we have set to 1. To first and second orders, namely, at two and three loops the diagram \nref{spherelong} contributes
\be\la{spherelongtwo}
- \({\lambda\over8\pi^2}\)^2\[8\pi{\log\delta\over\delta} +\dots\]\   -\ \({\lambda\over8\pi^2}\)^3\[{8\pi^4\over3}{1\over\delta^2} +\dots\]
\ee
to $\Gamma_{\rm cusp}$, in agreement with the leading terms in
(\ref{twoloops}) and (\ref{threeloops}). Here we have set $\epsilon_{UV} \sim \delta$ in
 \nref{spherelong}.
 Note, in particular, that the funny $1/\delta^2$ term
is coming from a power law IR divergence in the three loop computation of the quark anti-quark
potential in flat space. The $\log^2(\delta)$ term in the three loop result (\ref{threeloops}) can be
obtained from the exponentiation of the two loop single log term in \nref{spherelongtwo},
see \cite{Pineda:2007kz}.
If instead of expanding in $\lambda$ we do the integral in \nref{spherelong} and then
expand in $\delta$ for $\delta \ll \lambda$, then we get the expected behavior
\be\la{smalldelta2}
 { \lambda \over 2 \pi^2 } \int\limits_{\epsilon_{UV}}^\infty { d t \over 2 (\cosh t -1) } ( e^{ - t {\lambda\over4\pi\delta} } - 1)
={ 1 \over \delta }\[ {\lambda^2\over8\pi^3}\log\lambda +\dots\]
 \ee
 at two loops. In order to get the subleading coefficients represented by the dots one needs to be more careful about the matching of the effective theory in the IR with the full theory, see \cite{Pineda:2007kz} for a full discussion\footnote{The integral \nref{smalldelta2} has a
 logarithmic UV divergence, which is absorbed as part of the matching procedure \cite{Pineda:2007kz}.}. In the next section we will reproduce the two loop result of \cite{Pineda:2007kz} by resumming ladder diagrams. We also do a similar resummation for the three loop ladder result. However, in the full theory  the
non-ladder diagrams do contribute.  We leave the problem of performing the full
three loop resummation   for the future.

\subsection{Ladders diagrams and the quark anti-quark potential}

Only ladder diagrams contributed to the one and two loop results (\ref{oneloop}), (\ref{twoloops}) of section \ref{sec:amplitudes}. This means that a resummation    of ladder diagrams would correctly reproduce the two loop quark anti-quark potential. Moreover, in section \ref{sec:ladders} we considered the limit $i\theta\to\infty$ where only ladder diagrams survive. To compute the potential between the corresponding special quarks one has to re-sum these diagrams. Hence, in this section we will perform that resummation up to three loops.

The Bethe-Salpeter equation maps the sum of ladder diagrams to the ground state energy of the Schr\"odinger problem \cite{Erickson:1999qv}
\be\la{Schroedinger}
\[-\partial_x^2-{\hat\lambda\over4\pi^2(x^2+1)}\]\psi_n(x)=-{\Omega_n^2\over4}\psi_n(x)
\ee
as
\be
\Omega_0=-r \, V(\lambda)=\hat\lambda c_1+\hat\lambda^2 c_2+\hat\lambda^3 c_3+\dots
\ee
The computation of $\Omega_0$ also contains the IR effects we discussed above. Therefore, the perturbative expansion in $\hat\lambda$ for the Schr\"odinger problem (\ref{Schroedinger}) is a bit subtle. For example, if we applied the same iterative strategy we took in section \ref{perturbativesphere} when solving for the Schr\"odinger problem on the sphere (\ref{schpro}), we would get divergent integrals. Instead, we take a different strategy as we now explain.

We divide the $x$ axis into the large $x$ (IR) region and the small $x$ (UV) region. The two regions overlap at $x \sim1/\sqrt{\hat\lambda}$ where the potential and the energy are of the same order. In the IR region the equation can be solved straightforwardly. There, the solution up to two loops is a Bessel function. At three loops we also have to expand the potential to next order in
 the $1/x$ expansion.
 In the UV region, we start from a constant wave function and iterate the equation to the desired order. The energy is then obtained by demanding that the two solutions match in the overlap region $x\sim1/\sqrt{\hat\lambda}$. In that way we find
\ba\la{ladders}
c_1\!\!\!&=&\!\!\!{1\over4\pi}\nn\\
c_2\!\!\!&=&\!\!\!{1\over8\pi^3}\[\log{\hat\lambda\over2\pi}+\gamma_E-1\]\\
c_3\!\!\!&=&\!\!\!{1\over32\pi^5}\[\(\log{\hat\lambda\over2\pi}+\gamma_E\)\(\log{\hat\lambda\over2\pi}+\gamma_E+1\)-{7\over2}-{\pi^2\over12}\]\nn
\ea
where $\gamma_E$ is the Euler-Mascheroni constant. The two loop result $c_2$ as well as the leading log in $c_3$ agrees with the computation of \cite{Pineda:2007kz} using the effective field theory techniques discussed above. {The two loop quark anti-quark potential at finite $\theta$ is given by simply substituting $\hat\lambda\to\lambda(1+\cos\theta)/2$. At three loops, for the first time, also non ladder diagrams contribute and the corresponding result differs from $c_3$. We leave that computation to future work.}

One can also study the Schr\"odinger problem in the limit where $\hat\lambda\to\infty$. In that limit we get a huge potential well and to leading order the ground state energy is simply given by the value of the potential at $x=0$. That is
\be \la{laddqqbar}
\Omega_0= { \sqrt{\hat\lambda}\over\pi } +\cO(1)
\ee
The exact same value is obtained by taking the $i\theta\to\infty$ limit of the strong coupling result of \cite{Maldacena:1998im}. That is a special case of the match obtained in section \ref{strongcoupling}, in the limit $\phi \to \pi$.
 In \cite{Erickson:1999qv} the sum over ladder diagrams was compared to the strong computing
 computations in \cite{Maldacena:1998im,Rey:1998ik}  {\it for zero $\theta$} and it was found
 to disagree. However, if one
 takes a quark antiquark with a relative $\theta$ angle, which was also computed in  \cite{Maldacena:1998im},  and one takes the large $i \theta$ limit, then one finds that it
 matches precisely with \nref{laddqqbar}.

 At small $\hat\lambda$, the potential in (\ref{Schroedinger}) has a single bound state. As we increase $\hat\lambda$ more and more bound states go down from the continuum. Beyond a critical value $\hat\lambda>\hat\lambda_c=\pi^2$ we have an infinite number of bound states \cite{Klebanov:2006jj}.
\footnote{In \cite{Klebanov:2006jj} it was also argued that this should be the behavior in the full theory
 for the antiparallel lines in flat space.

 }
One may wonder whether also these can be matched with the modes of the string at strong coupling. We find that unlikely as generically the string has infinitely many degrees of freedom and not just one.
Note however that the density of states near the top of the potential at $\Omega_\infty=0$ does match with the string density of states for the relevant mode \cite{Klebanov:2006jj}. This
can be interpreted as the statement that the classical motion for the string mode identified
in \cite{Klebanov:2006jj} matches the classical   (or large $\hat \lambda$) limit of the Schr\"odinger problem.

\section{Conclusions}

In this paper we have performed a three loop computation of the cusp anomalous dimension
for the locally supersymmetric Wilson loop in ${\cal N}=4$ super Yang Mills. Equivalently, we
have computed the three loop quark/anti-quark potential on the three sphere.
The final result is eq. \nref{gamma3_intro}.
The computation was done by considering the scattering of massive $W$ bosons and
focusing on the IR divergences, which are given by $\Gamma_{cusp}(\varphi)$. In turn,
this amplitude is related by dual conformal symmetry to the Regge limit of the four
point amplitude of massless particles in the Higgs regularization \cite{Alday:2009zm,Henn:2010bk}.
We have discussed some checks of this result.
In particular, we have matched the leading $\phi^2$ term to the exact computation in
\cite{Correa:2012at,Fiol:2012sg}. We have also discussed the limit that corresponds
to the quark anti-quark potential. In this limit, small $\delta$,
 we get a result that is different from
the $1/\delta$ behavior expected naively, where $\delta$ is the relative angle. This is
due to some divergences that appear in perturbation theory in this computation
\cite{Appelquist:1977es,Erickson:1999qv,Pineda:2007kz}. After taking into account the
long distance effective theory describing these IR effects one can explain the behavior
of the result for small $\delta$. In principle, one should be able to do a resummation
and give the full three loop result for the flat space quark-antiquark potential.
This is a problem we left to the future.

We have also identified a large $i \theta$ limit \nref{scalinglimit}
of the generalized cusp anomalous dimension,
$\Gamma_{cusp}(\phi,\theta)$,  where only ladder diagrams contribute. These ladder diagrams
can be summed by solving a simple Schr\"odinger problem. For general angles the potential
does not appear to be solvable, but one can develop a simple perturbative expansion where
one clearly sees that the answer is given by iterated integrals that give Goncharov polylogarithms.
 For a particular angle, $\phi =0$, we could solve the problem and sum
the ladder diagrams for all effective couplings $\hat \lambda$. It is also possible to
get the first term in the $\phi^2$ expansion around $\phi=0$. The final answer is in \nref{full_phi2}. Also, we have taken the small $\delta$ limit of the Schr\"odinger problem and
expanded the answer to third order in $\hat \lambda$. We reproduced the two loop resummed answer in \cite{Pineda:2007kz}.
It is very simple to compute the strong coupling limit of the result for the ladder diagrams. More
precisely, we take the effective coupling $\hat \lambda$ to be very large. Interestingly
we find a precise match with strong coupling string theory computations. Here one first
takes $\lambda$ to be large, and then takes $e^{ i\theta } \to \infty$. In this limit
the classical string theory answers go like $e^{ i\theta/2}$ so that multiplied by
$\sqrt{\lambda}$ one gets  $\sqrt{\hat \lambda}$ which is the behavior of the strong
coupling ladder diagrams. This agreement was not preordained, since the scaling limit
\nref{scalinglimit} is explicitly pushing us towards the small $\lambda$ region.
It seems that the origin of this agreement has the same underlying reason as the agreement
for the BMN \cite{Berenstein:2002jq} or large charge limits \cite{Kruczenski:2003gt}.

{\bf Acknowledgements }

We would like to thank N. Arkani-Hamed, S. Caron-Huot and N. Gromov for discussions.
We thank T. Riemann for providing a Mathematica file of the integrals computed in ref. \cite{Czakon:2004wm}.
We thank T. Huber for pointing out typos in the first version of this paper.

This work was supported in part by   U.S.~Department of Energy grant \#DE-FG02-90ER40542.
Research at the Perimeter Institute is supported in part by the Government of Canada through NSERC and by the Province of Ontario through MRI.  The research of  A.S. has been supported in part by the Province of Ontario through ERA grant ER 06-02-293. D.C would like to thanks IAS for hospitality.
The research of D.C has been supported in part by a CONICET-Fulbright fellowship and grant PICT 2010-0724.

\appendix

\section{Results for Wilson loop integrals relevant for QCD}\la{QCDintegrals}

It is important to realize that, at weak coupling, Wilson loops depend only mildly on the
particle content. There are two sources of differences between $\cN=4$ SYM and QCD
for these Wilson loops. The first is the additional coupling of the loop to scalars in $\cN=4$ SYM.
The second is the specific particle content, which e.g. enters at two loops through the one-loop
gluon propagator correction. We want to stress   that many diagrams that have to be
calculated are identical. This means that our result computes a part of the QCD result.

Moreover, there is a direct relation between some of the Wilson loop integrals
we are interested in, and the infrared divergent part of form factor integrals
discussed in the literature.
These connections are very helpful. First of all, they give an idea about the kind
of functions that can appear in such calculations.
In some cases the relations are even more concrete,
and they provide two different ways of computing the same object.
We will give two examples.
The first one relates the $\xi^2$ term in $\Gamma_{\rm cusp}^{(2)}$
to a known form factor integral. Inversely, we point out that our analytic
result at three loops, obtained from scattering amplitudes, implies that
some three-loop Wilson line integrals that are relevant for QCD are now known.
In the second example, we discuss the horizontal ladder diagram, which contributes to
the $\xi$ term. This is just one contribution to the $\xi$ term but there are other
 contributions which are not these ladder diagrams.
 Here, the Wilson loop picture turns out to be extremely simple,
and allows us to derive an all-orders result for this class of integrals.
Of course, to get the QCD answers we need to set $\xi = { \cos \phi \over \sin \phi}$.

\subsection{Results for crossed ladder integrals at two and three loops}

It is easy to see from the Wilson loop computation that the term proportional
to $\xi^2$ at two loops comes from the ladder with two rungs and a contribution
from the squared one-loop ladder.
 It is well known that this combination
results in a crossed ladder Wilson line integral, see Fig.~\ref{fig:crossed-ladders}(a).
The relevant UV divergent part of that integral can be equivalently obtained from
the IR pole of the form factor integral of the same topology. The latter is given by
$V_{6l4m1;{-1}}$ in ref. \cite{Czakon:2004wm}. Comparing to eq. (\ref{gammacusp2loop}),
we find perfect agreement.

\begin{figure}[t]
\begin{center}
\includegraphics[width=1.0\textwidth]{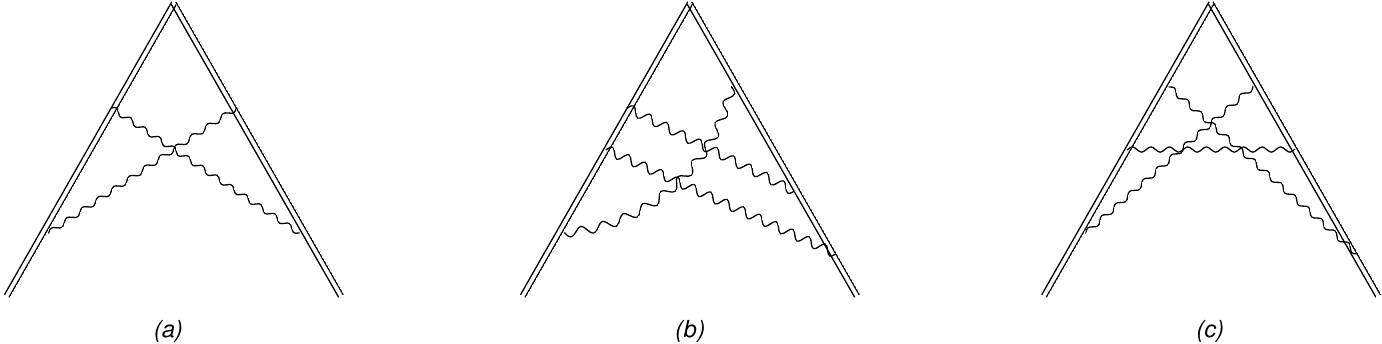}
\caption{Crossed ladder diagrams that  appear in the computation of $\Gamma_{\rm cusp}$
at two and three loops.}
\label{fig:crossed-ladders}
\end{center}
\end{figure}

Having obtained the full three-loop result for $\Gamma_{\rm cusp}$ by evaluating the
 four-point amplitude in the Regge limit as described in section \ref{sec:amplitudes},
we can provide an analytical answer for some of the Wilson loop diagrams, and equivalently,
form factor integrals that are of more general interest. In fact, the $\xi^3$ terms in $\Gamma_{\rm cusp}^{(3)}$
come from summing ladder diagrams only. Taking into account the exponentiation of lower-loop
graphs, they correspond to the sum of the Wilson loop diagrams shown in Fig.~\ref{fig:crossed-ladders}(b,c).
Therefore, we have that
\begin{align}
\left[ {\rm Fig.}\, \ref{fig:crossed-ladders}(b) +  {\rm Fig.}\,\ref{fig:crossed-ladders}(c)   \right] \, \sim \, \log \Lambda \, \times  \Big[ \xi^3 {\;term\; in\; eq. \; (\ref{gamma3_intro})} \Big]  + \cO(\Lambda^0 )\,,
\end{align}
where $\Lambda$ is some UV cutoff. Similarly, in dimensional regularization $\log \Lambda$ would be replaced by $1/\eps$.
Being built from pure gluon propagators, these diagrams also appear e.g. in the QCD calculation of $\Gamma_{\rm cusp}$ at three loops,
or equivalently when computing massive form factors at three loops.

\subsection{Exact result for a class of diagrams}

\begin{figure}[t]
\begin{center}
\includegraphics[width=0.95\textwidth]{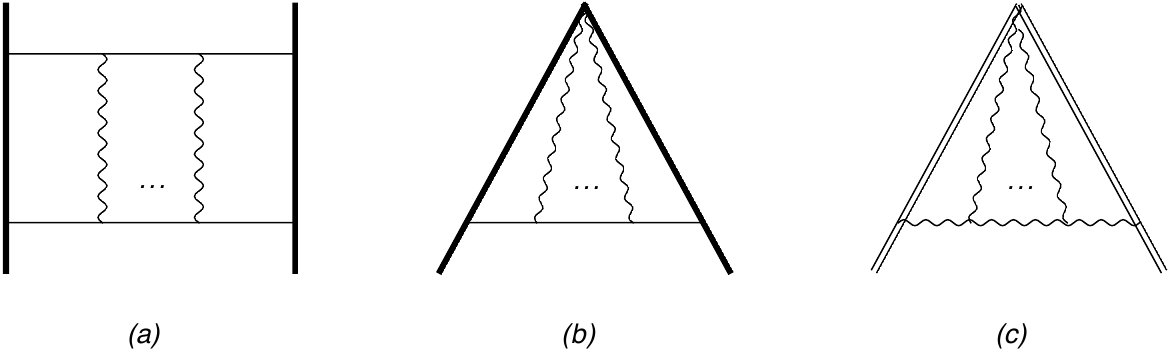}
\caption{Horizontal ladder diagram (a) in the soft limit $m^2/s \to 0$. Thick solid lines denote heavy particles
of mass $M$, thin solid lines denote particles of mass $m$, and wavy lines denote massless particles. The leading IR divergence
of (a) is given by twice or IR divergence of the vertex integral (b). The divergent part of the latter is captured by the eikonal limit, leading
to the Wilson line integral (c). The doubled lines in the latter denote the Wilson line contour. At $L>1$ loops,
the massless coordinate space integral in the interior can be identified with the off-shell ladder integral $\Phi^{(L-1)}$.}
\label{fig:horizontal-ladders}
\end{center}
\end{figure}

Here we compute the Regge limit of the horizontal ladder diagrams, see Fig.~\ref{fig:horizontal-ladders}(a), to all loop orders,
by relating them to known off-shell ladder diagrams \cite{Usyukina:1993ch}. We do not know
if there is a physical limit that isolates only these diagrams.
For massive scalar integrals, there is a useful counting procedure to determine their rate of divergence in the Regge limit \cite{Eden}.
Applied to the present case, we find the following behavior,
\begin{align}
I^{(L)}_{\rm horizontal \; ladder}(u,v)   \underset{u\to 0}{\longrightarrow}  \log u \,\, R^{(L)}(\varphi) + \cO(u^0) \,.
\end{align}
We want to compute the function $ R^{(L)}(\varphi)$ that multiplies the logarithm.
It is well known that there are simplifications for that term at the
level of Feynman parameter integrals \cite{Eden}.

Here, we follow a different approach.
We find it more convenient to think of the limit $u\to0$ as the soft limit
in the Bhabha scattering picture, as discussed above.
At $L$ loops, it is clear that the leading soft divergence comes from regions where all loop momenta are
soft. There is a factor of $2$ because there are two relevant regions.
In this limit, it is easy to see that the ladder becomes a three-point diagram with eikonal lines, cf. Fig.~\ref{fig:horizontal-ladders}(b).
The latter can be represented as a double line integral, cf. Fig.~\ref{fig:horizontal-ladders}(c), and we obtain a result proportional to (for $L>1$),
\begin{align}\label{IRUVcancellation}
I^{(L)}_{\rm horizontal \; ladder} \underset{\rm eikonal}{\sim} \int\limits_{0}^{\infty} d\tau_{1} d\tau_{2} \frac{(p_{2}+p_{3})^2}{y_{12}^2}  \Phi^{(L-1)}\left( \frac{y_{1}^2}{y_{12}^2}, \frac{y_{2}^2}{y_{12}^2} \right) \,,
\end{align}
where
\begin{align}
y^{\mu}_{1}(\tau_{1}) = \tau_1 p_{2} ^{\mu} \,,\qquad  y^{\mu}_{2}(\tau_{2}) = \tau_{2} p_{3} ^{\mu} \,, \qquad y_{12} = y_1 - y_2 \,,
\end{align}
and where $\Phi^{(L)}$ are known off-shell ladder integrals.
They are given by a linear combination of logarithms and polylogarithms of homogeneous degree $2 L$ \cite{Usyukina:1993ch}.

By taking the eikonal limit, we have in fact introduced UV divergences for small $\tau_{1,2}$, and as a result,
the integral in eq. (\ref{IRUVcancellation}) is formally zero. The coefficient of the IR logarithm we want to compute
is given by minus the coefficient of the UV logarithm.
The latter can be easily extracted from the small $\tau_{1,2}$ region of the integral,
by introducing a cut-off for the radial integration.
Collecting all proportionality factors, we find
\begin{align}\label{turnedladders2}
R^{(L)}(\varphi) = -2 \frac{\cosh \varphi -1}{\sinh \varphi} \int\limits_{0}^{1} \frac{d z}{z} \, \tilde{\Phi}^{(L-1)}\left( \frac{1}{1+2 z \cosh \varphi + z^2},\frac{z^2}{1+2 z \cosh \varphi + z^2} \right) \,,
\end{align}
where ${\tilde\Phi}$ are rescaled off-shell ladder integrals,
\begin{align}
\tilde{\Phi}^{(L)}(x,y) = \lambda(x,y) \Phi^{(L)}(x,y)\,,\qquad \lambda(x,y) = \sqrt{(1-x-y)^2-4x y}\,.
\end{align}
For $L=2$, eq. (\ref{turnedladders2}) has appeared in the Wilson loop computation of ref. \cite{Makeenko:2006ds} and was evaluated
analytically in ref. \cite{Drukker:2011za}. Here we show how to perform the integration for any $L$.

The integral can be easily evaluated by using the following integral representation for  $\Phi^{(L)}$ \cite{Usyukina:1993ch},
valid for $x>y$,
\begin{align}
\Phi^{(L)}(x,y) = -\frac{1}{L! (L-1)!} \int\limits_{0}^{1} \frac{dt}{y t^2+(1-x-y) t+x}  \left[ \log t   \log \left(\frac{y }{x}  t \right) \right]^{L-1}   \log\left( \frac{y}{x} t^2 \right) \,.
\end{align}
Plugging this formula into eq. (\ref{turnedladders2}), we notice that the change of
variables $t = \rho \, e^{-\tau}, z=\rho\, e^{\tau}$ makes one
integration trivial (after expanding powers of logarithms into sums),
while the remaining one reduces to a known integral\footnote{Gradshteyn and Ryzhik, ``Table of Integrals, Series, and Products'', Academic Press, Sixth Edition, p.~549, equation (14).}.
In this way, we arrive at the result
\begin{align}\label{turnedladders3}
R^{(L+1)}(\varphi) = \,  \xi \,  g(L) \,  \left[ \pi \, \left(\frac{\partial}{\partial{x}}\right)^{2 L} \frac{\sinh(x\,  \varphi)}{\sin(x\, \pi)} \right]_{x=0}\,.
\end{align}
Here $g(L)$ is given by a sum,
\begin{align} \la{gval}
g(L) =  - \frac{{2^{-2 L+3}}}{L! (L-1)!}   \sum_{m_1,m_2 = 0}^{L-1} \binom{L-1}{m_1} \binom{L-1}{m_2} \frac{3^{L-1-m_2}}{1+m_1+m_2} \left( (-1)^{m_1} + (-1)^{m_2} \right) \,,
\end{align}
and hence it evaluates to rational numbers at any $L$. Inspecting eq. (\ref{turnedladders3}), \nref{gval}
 we see that
\begin{align}\label{turnedladders4}
R^{(L+1)}(\varphi) = \,  \xi \,  \frac{(-1)^{L+1} (4\pi)^{2L+1}}{(2L+1)!} i B_{2L+1}\left(\frac{\pi-i\varphi}{2\pi}\right)
\,.
\end{align}
where $B_n(x)$ is the Bernoulli polynomial. Up to the $\xi$ factor, the answer for $R^{(L)}$ is a polynomial in $\pi$ and $\varphi$! The polynomial is homogeneous, and of degree $2L-1$.

For example, at two and three loops, we immediately reproduce eqs.
(\ref{ladderor1loop2}) and (\ref{ladderor1loop3}), obtained by an independent computation using Mellin-Barnes methods.
At four loops, we obtain
\begin{align}\label{horizontalladder4loops}
R^{(4)}(\varphi) = - \xi  \frac{8}{945} \left(  31\, \varphi\, \pi^6 + 49 \, \varphi^3 \, \pi^4 + 21 \, \varphi^5 \, \pi^2+  3\, \varphi^7 \right) \,.
\end{align}
Let us perform a consistency check on this result
by taking the limit $x \to 0$. In the latter, $\xi \to 1$, and $\varphi = - \log x$ stays invariant.
We can see that this is consistent with eq. (A.1) of ref. \cite{Henn:2010ir}, where this limit was
previously computed.

At $L$ loops, we know from unitarity cuts that these ladders appear with coefficient $(-\lambda/(16 \pi^2))^L$ in
the amplitude.
Therefore, they give the following contribution to $\Gamma_{cusp}$
\begin{align}
\Gamma_{\rm cusp}^{\rm (horizontal\; ladders)} =&  \sum_{L\ge 1}  \left( \frac{-\lambda}{16 \pi^2} \right)^L  \;R^{(L)}(\varphi)  \nonumber \\
=& \frac{i \,\xi}{4 \pi  }  \,  \sum_{L\ge 1}  \, \frac{ \lambda^L }{(2L-1)!} B_{2L-1}\left(\frac{\pi-i\varphi}{2\pi}\right)
\,.
\end{align}

In summary, we have computed the leading Regge limit of the horizontal ladder diagram shown in Fig.~\ref{fig:horizontal-ladders}(a),
or equivalently, the coefficient of the pole term in the corresponding massive form factor integral shown in Fig.~\ref{fig:horizontal-ladders}(b),
at any loop order $L$. Up to an overall factor of $\xi = (\cosh \varphi - 1)/\sinh \varphi$, the result is a homogeneous polynomial of degree $2L-1$ in $\varphi$ and $\pi$.

Interestingly, the horizontal ladders result (\ref{turnedladders4}) also appears in the TBA
integrability approach to the cusp anomalous dimension \cite{IntegrabilityPaper}. There, it appears as
the leading L\"uscher contribution with $L+1$ scalar insertions at the cusp.

The computations are superficially different.
Let us now give an argument for this agreement\footnote{They agree up to a minus sign.}.
The calculation of the L\"uscher contribution would be the following.
Considers a Wilson loop with two cusps, and with $Z^L$ and $\bar{Z}^L$ inserted
at the two cusps, respectively. Here $Z = \Phi_{5} + i \Phi_{6}$ are complex scalar fields.
The Wilson lines connecting the cusps are smooth
and couple to scalars different from $\Phi_{5}$ and $\Phi_{6}$.
At tree level, one only has Wick contractions between the $Z$ and $\bar{Z}$ fields.
We are interested in the first wrapping-type diagrams, where a field is emitted
from one Wilson line, interacts with the $Z$-$\bar{Z}$ propagators, and is absorbed
by the other Wilson line. This term will be linear in $\xi$. We will focus on the
term where the scalars are emitted and absorbed.
We now argue that this calculation would lead to a result proportional to (\ref{turnedladders4}).

The argument consists of two steps. First, notice that there is only one
type of Feynman diagram we need to consider. The reason is that the $SU(4)$ charges of the scalars have to flow through the diagram.
The only way to connect the diagram in a planar way is to use four-scalar vertices.
We are interested in the scaling dimension of the cusp with operator insertion,
so we need to compute only the UV divergences of this diagram.
The latter come from the integration regions close to the cusps.
We can focus on one cusp to compute them. When doing so, at leading order
the propagators connecting to the other cusp factor out, and we are left with the diagram
of figure \ref{fig:horizontal-ladders}(c).
This shows that the two calculations have to agree, up to the
overall coefficient.

One may wonder whether there is a situation where only the infinite class of integrals computed here
contributes to a physical quantity. In section \ref{sec:ladders}, we saw that this was the case for a similar
class
of ladder diagrams, which could be singled out by taking the large $i \theta$ limit, where $\theta$
determines
how the Wilson lines couple to the scalars. More generally, we can think of the scattering amplitude
depending
on several angles on $S^{5}$, as explained in the introduction.
Then, the limit of section \ref{sec:ladders} just selects the vertical ladder diagrams. It seems possible that
 a similar limit selects the horizontal ladder diagrams.

\section{Results for individual integrals up to three loops}
\label{appendix-integrals}

Here we present analytic formulas for the integrals contributing to
$\Gamma_{\rm cusp}$ to three loops.
The Regge limit of the one-loop integral was already given
in the main text.
At two loops, we find
\begin{align}
I_{2,1} =& \, \xi \, \left[  \frac{4}{3} \pi^2  \log x + \frac{4}{3} \log^3 x  \right]  \,, \label{ladderor1loop2} \\
I_{2,0} =& \, \xi \, \Big[ -16 H_{-3,0}(x)-16 H_{3,0}(x)-16 H_{2,0,0}(x)-8 H_{0,0,0,0}(x) \nonumber \\
&\quad\quad -24 \zeta _2 H_{0,0}(x)  -4
   \zeta _3 H_0(x)-42 \zeta _4 \Big] \,, \\
I^{r}_{2,2}=& \, \xi^2 \, \left[ 2 \log^2 x \right]  \,, \\
I^{r}_{2,1} =& \, \xi^2 \, \Big[ -16 \zeta_2 H_{0}(x) - 16 H_{2,0}(x)-16 H_{-1,0,0}(x)-8 H_{0,0,0}(x)-16 H_{1,0,0}(x) - 4 \zeta_3 \Big]\,, \\
I^{r}_{2,0} =& \, \xi^2 \, \Big[
32 \zeta _2 H_{-1,0}(x)+24 \zeta _2 H_{0,0}(x)+32 \zeta _2 H_{1,0}(x)+16 H_{-3,0}(x)+16 H_{3,0}(x) \nonumber \\
& \quad\quad -32 H_{-2,-1,0}(x)+16 H_{-2,0,0}(x)+32 H_{-1,2,0}(x)+32 H_{1,2,0}(x)+16 H_{2,0,0}(x) \nonumber \\
& \quad\quad +32 H_{2,1,0}(x)+32 H_{-1,-1,0,0}(x)+16 H_{-1,0,0,0}(x)+32 H_{-1,1,0,0}(x)+8 H_{0,0,0,0}(x) \nonumber \\
& \quad\quad +32 H_{1,-1,0,0}(x)+16 H_{1,0,0,0}(x)+32 H_{1,1,0,0}(x) \nonumber \\
& \quad\quad -16 \zeta _2 H_{-2}(x)+8 \zeta _3 H_{-1}(x)+8 \zeta _3 H_1(x)+32 \zeta
   _2 H_2(x)+60 \zeta _4 \,.
 \Big] \,.
\end{align}
At three loops, we have
\begin{align}
I_{3a}(u,v)  \,\; \underset{u\to 0}{\longrightarrow} \,\; & \log u \, I_{3a,1} + \cO(u^0) \,, \\
I_{3a}(v,u)  \,\; \underset{u\to 0}{\longrightarrow} \,\; & \log^3 u \, I^{r}_{3a,3}+\log^2 u \, I^{r}_{3a,2}+\log u \, I^{r}_{3a,1} + \cO(u^0) \,, \\
I_{3b}(u,v)  \,\; \underset{u\to 0}{\longrightarrow} \,\; & \log^2 u \, I_{3b,2} + \log u \, I_{3b,1} + \cO(u^0) \,, \\
I_{3b}(v,u)  \,\; \underset{u\to 0}{\longrightarrow} \,\; & \log u \, I^{r}_{3b,1} + \cO(u^0) \,.
\end{align}
We found the following coefficient functions,
\begin{align}\label{ladderor1loop3}
I_{3a,1} =&  \, \xi  \,  \frac{4}{45}  \left[7 \pi^4 \log x  + 10 \pi^2 \log^3 x  + 3 \log^5 x \right]  \,,
\end{align}
which is in agreement with the calculation for general $L$, c.f. eq. (\ref{turnedladders3}),
{
\allowdisplaybreaks
\begin{align}
I^{r}_{3a,3} =& \frac{1}{6} (I_{1,1})^3 \,,\\
I^{r}_{3a,2} =&  \,
\xi^3  \,   \Big[
-40 \zeta _2 H_{0,0}(x)+16 H_{-3,0}(x)-48 H_{3,0}(x)-64 H_{2,0,0}(x) \nonumber \\
&\quad \quad -48
   H_{-1,0,0,0}(x)-64 H_{0,0,0,0}(x)-48 H_{1,0,0,0}(x)-8 \zeta _3 H_0(x)
 \Big] \,,  \\
I^{r}_{3a,1} =& \xi^3 \, \Big[
32 \zeta _3 H_{-1,0}(x)+16 \zeta _3 H_{0,0}(x)+32 \zeta _3 H_{1,0}(x)-32 \zeta _2
   H_{-2,0}(x)+224 \zeta _2 H_{2,0}(x) \nonumber \\
   & \vspace{-0.5cm} +160 \zeta _2 H_{-1,0,0}(x)+256 \zeta _2
   H_{0,0,0}(x)+160 \zeta _2 H_{1,0,0}(x)+256 H_{4,0}(x)-64 H_{-3,-1,0}(x)  \nonumber \\
   & \vspace{-0.5cm} +128
   H_{-3,0,0}(x)-64 H_{-3,1,0}(x)-64 H_{-2,-2,0}(x)-64 H_{-1,-3,0}(x)+192
   H_{-1,3,0}(x) \nonumber \\
   & \vspace{-0.5cm} -64 H_{1,-3,0}(x)+192 H_{1,3,0}(x)-64 H_{2,-2,0}(x)+256 H_{2,2,0}(x)-64
   H_{3,-1,0}(x) \nonumber \\
   & \vspace{-0.5cm} +256 H_{3,0,0}(x)+192 H_{3,1,0}(x)-192 H_{-2,-1,0,0}(x)+160
   H_{-2,0,0,0}(x)+64 H_{-2,1,0,0}(x) \nonumber \\
   & \vspace{-0.5cm} +256 H_{-1,2,0,0}(x)+256 H_{1,2,0,0}(x)+64
   H_{2,-1,0,0}(x)+288 H_{2,0,0,0}(x)+320 H_{2,1,0,0}(x) \nonumber \\
   & \vspace{-0.5cm} +192 H_{-1,-1,0,0,0}(x)+256
   H_{-1,0,0,0,0}(x)+192 H_{-1,1,0,0,0}(x)+272 H_{0,0,0,0,0}(x) \nonumber \\
   & \vspace{-0.5cm} +192
   H_{1,-1,0,0,0}(x)+256 H_{1,0,0,0,0}(x)+192 H_{1,1,0,0,0}(x)-16 \zeta _3 H_{-2}(x)+48
   \zeta _3 H_2(x)  \nonumber \\
   & \vspace{-0.5cm}  -96 \zeta _2 H_{-3}(x)+204 \zeta _4 H_0(x)+160 \zeta _2 H_3(x)+32
   \zeta _2 \zeta _3+12 \zeta _5
 \Big]     \,,
\end{align}
}
and for $I_{3b}(u,v)$:
\begin{align}
I_{3b,2} \, =& + \frac{1}{2} I_{1,1} I_{2,1} \,, \\
I_{3b,1} \, =& \xi^2 \,  \Big[\,
-128 H_{4,0}(x)-64 H_{-3,0,0}(x)-96 H_{3,0,0}(x)-144 H_{2,0,0,0}(x)-128 H_{1,0,0,0,0}(x) \nonumber \\
   & \quad\quad -216
   H_{0,0,0,0,0}(x)  -32 \zeta _2 H_{2,0}(x)-176 \zeta _2 H_{0,0,0}(x)-64 \zeta _2
   H_{1,0,0}(x) \nonumber \\
      & \quad\quad    -16 \zeta _3 H_{0,0}(x) -152 \zeta _4 H_0(x)-24 \zeta _5
  \Big]\,.
\end{align}
Finally, we have
\begin{align}
I^{r}_{3b,1} \, =&  \xi \, \Big[
\frac{8}{15} \log ^5 x+\frac{8}{9} \pi ^2 \log ^3 x+\frac{16}{45} \pi ^4 \log x
\Big] \,.
\end{align}
We remark that the factorization of the leading terms $I^{r}_{2,2},
I^{r}_{3a,3},  I_{3b,2}$ can be understood from the analysis of
\cite{Eden,Henn:2010ir}, where the systematics of the Regge limit
were investigated.

\end{document}

%% file: fig1.pdf_tex
\begingroup%
  \makeatletter%
  \providecommand\color[2][]{%
    \errmessage{(Inkscape) Color is used for the text in Inkscape, but the package 'color.sty' is not loaded}%
    \renewcommand\color[2][]{}%
  }%
  \providecommand\transparent[1]{%
    \errmessage{(Inkscape) Transparency is used (non-zero) for the text in Inkscape, but the package 'transparent.sty' is not loaded}%
    \renewcommand\transparent[1]{}%
  }%
  \providecommand\rotatebox[2]{#2}%
  \ifx\svgwidth\undefined%
    \setlength{\unitlength}{806.43196593bp}%
    \ifx\svgscale\undefined%
      \relax%
    \else%
      \setlength{\unitlength}{\unitlength * \real{\svgscale}}%
    \fi%
  \else%
    \setlength{\unitlength}{\svgwidth}%
  \fi%
  \global\let\svgwidth\undefined%
  \global\let\svgscale\undefined%
  \makeatother%
  \begin{picture}(1,0.43943814)%
    \put(0,0){\includegraphics[width=\unitlength]{fig1.pdf}}%
    \put(0.48807166,0.10294177){\color[rgb]{0,0,0}\makebox(0,0)[lb]{\smash{$\phi$}}}%
    \put(0.86412969,0.09094119){\color[rgb]{0,0,0}\makebox(0,0)[lb]{\smash{$\phi$}}}%
    \put(0.25921539,0.00430436){\color[rgb]{0,0,0}\makebox(0,0)[lb]{\smash{$({\bf a})$}}}%
    \put(0.7949084,0.00430436){\color[rgb]{0,0,0}\makebox(0,0)[lb]{\smash{$({\bf b})$}}}%
    \put(0.96227488,0.08876836){\color[rgb]{0,0,0}\makebox(0,0)[lb]{\smash{$S^3$}}}%
    \put(0.08131177,0.13888338){\color[rgb]{0,0,0}\makebox(0,0)[lb]{\smash{$\Blue{\vec n}$}}}%
    \put(0.41289867,0.17514362){\color[rgb]{0,0,0}\makebox(0,0)[lb]{\smash{$\Blue{\vec n'}$}}}%
    \put(0.80659229,0.07308474){\color[rgb]{0,0,0}\makebox(0,0)[lb]{\smash{$\delta$}}}%
  \end{picture}%
\endgroup%